\documentclass[a4wide,twoside]{article}
\usepackage{latexsym,epsfig}
\usepackage{fancyheadings}
\usepackage[english]{babel}
\usepackage{amsmath,amssymb,amsfonts}
\usepackage[footnotesize,bf]{caption}
\usepackage{enumerate}
\usepackage{array,dcolumn}
\usepackage{graphicx}
\usepackage{psfrag}

\def\Mpi{M_{\pi}^2}
\def\Mpin{M_{\pi^0}^2}
\def\Mpic{M_{\pi^{\pm}}^2}
\def\mupi{\mu_{\pi}}
\def\mupin{\mu_{\pi^0}}
\def\mupic{\mu_{\pi^{\pm}}}
\def\mupitild{\tilde{\mu}_{\pi}}

\def\MK{M_{K}^2}
\def\MKn{M_{K^0}^2}
\def\MKc{M_{K^{\pm}}^2}
\def\muK{\mu_{K}}
\def\muKn{\mu_{{K}^0}}
\def\muKc{\mu_{{K}^{\pm}}}
\def\muKtild{\tilde{\mu}_{{K}}}

\def\Meta{M_{\eta}}
\def\mueta{\mu_{\eta}}

{\catcode `\@=11 \global\let\AddtoReset=\@addtoreset}
\AddtoReset{equation}{section}
\renewcommand{\theequation}{\thesection.\arabic{equation}}  

\newcommand{\lapprox}{%
\mathrel{%
\setbox0=\hbox{$<$}
\raise0.6ex\copy0\kern-\wd0
\lower0.65ex\hbox{$\sim$}
}}

\newcommand{\beq}{\begin{equation}}
\newcommand{\eeq}{\end{equation}}
\newcommand{\beqa}{\begin{eqnarray}}
\newcommand{\eeqa}{\end{eqnarray}}
\newcommand{\beqan}{\begin{eqnarray*}}
\newcommand{\eeqan}{\end{eqnarray*}}
\newcommand{\ba}{\begin{array}}
\newcommand{\ea}{\end{array}}
 
\newcommand{\bea}{\begin{eqnarray}}
\newcommand{\eea}{\end{eqnarray}}
\newcommand{\no}{\nonumber}

\normalsize
\begin{document}
\allowdisplaybreaks
\bibliographystyle{plain}
\begin{titlepage}
\begin{flushright}
CPT-2001/P.4228\\
\end{flushright}
\vspace{2.5cm}
\begin{center}
{\Large 
\bf Isospin Breaking Corrections to Low-Energy $\pi - K$ Scattering}\\[40pt]
A. Nehme\footnote{nehme@cpt.univ-mrs.fr}  and
P. Talavera

\vspace{1cm}

 {\it Centre de Physique Th\'eorique, CNRS--Luminy, Case 907\\
 F-13288 Marseille Cedex 9, France.}
\\[10pt]

\vfill
{\bf Abstract} \\
\end{center}
We evaluate the matrix elements 
for the processes
$\pi^0 K^0 \rightarrow \pi^0 K^0$  and
$\pi^- K^+ \rightarrow \pi^0 K^0$  
in the presence of isospin breaking terms at leading and next-to-leading
order. As a direct application the relevant combination of the 
S-wave scattering lengths involved in the pion-kaon atom lifetime is determined.
We discuss the sensitivity of the 
results with respect to the input parameters. \\


\vfill

\end{titlepage}

\setcounter{page}{0}

\clearpage
\tableofcontents

\addtocounter{page}{1}
\setcounter{equation}{0}
\section{Introduction}
\label{sec: Introduction}
\renewcommand{\theequation}{\arabic{section}.\arabic{equation}}
\setcounter{equation}{0}

Chiral perturbation theory \cite{Weinberg79} has become one of the most 
used tools in exploring QCD low-energy dynamics. It applies in the 
non-perturbative 
regime of QCD. 
The initial QCD lagrangean, ${\cal L}_0$, 
is then replaced by an
effective one which contains the same 
symmetries as the fundamental ${\cal L}_0$, 
composed by a string of higher and higher 
dimension operators involving derivatives and quark masses. 
Technically, the new lagrangean is not renormalizable
(in the Wilsonian sense) but     
fortunately at a given order in the momenta (quark mass) expansion 
the number of needed counter-terms is finite. Thus \emph{assuming} that
the chiral series converges \cite{dyson} one can truncate it at a given order and deal
with a finite number of unknown constants. Restricting the analysis to the 
next-to-leading order in the mesonic sector (see below) one can obtain 
the unknown constants
from the existing data and large-N$_c$ arguments. To this respect
one makes use of the experimental knowledge on the 
pseudoscalar masses and decay constants, pion vector form-factor  
and
K$_{\ell 4}$ form-factors. Therefore none of those data 
informations can be used to claim any theoretical \emph{predictability}. 
To exhibit the consistency of the theory  
one has to use other processes where the low-energy 
constants are given as mere inputs and confront the 
theoretical
results with the experimental data. 
To this aim the $\pi-\pi$ scattering lengths have deserved a careful
examination \cite{saga} but unfortunately they only bring information about
the SU(2) sector. In line with the previous general argument, 
$\pi-K$ scattering 
stems for the simplest meson-meson scattering process that involves
strangeness 
and can be used as an independent
source of information on the validity of the extra assumptions that common 
wisdom
assesses to hold in chiral perturbation theory, 
as for instance large-N$_c$ arguments.
This will, hopefully, bring some insight on the role of the strange 
quark mass inside the chiral expansion.  
Recently it has been noticed that some observables depend
strongly on the number of light sea quarks \cite{bachir}.  
This fact can cast doubts on the validity of the chiral expansion
in the SU(3) sector where
m$_s$ is treated as a small parameter.
Before any judgment is taken
it is necessary to make more accurate experiments and precision calculations
on \emph{testing processes}. In that sense the next proposal for 
the measurement of the lifetime and lamb-shift in $\pi-K$ atoms 
($A_{\pi K}$) at CERN 
constitutes a major step from the experimental side \cite{proposal}. 
Experiments on this 
system will constitute one of the most stringent test on chiral symmetry
breaking existing up to the moment. Even though we
want to stress that our treatment will not allow to deal with bound states, and a more
refined analysis in the line of the one performed for the $\pi-\pi$ atom is needed
\cite{harup,bernisses}. In a very brief way
the lifetime ($\tau$)  
of the A$_{\pi K}$ atom,
is given in terms of \cite{transp}
\[
\frac{1}{\tau_{n,0}}\propto 
(a_0^{1/2}-a_0^{3/2})^2\,. 
\]
Thus any theoretical insight on the shift for the scattering lengths due to the
isospin breaking terms constitutes a key role for a real estimate of A$_{\pi K}$ atom
lifetime. 

In this paper our aim is
to incorporate some of the theoretical effects that were not
taken into account in a previous analysis \cite{BKM1}. 
We shall deal first with the
more \emph{academic} $\pi^0 K^0 \rightarrow \pi^0 K^0$ process where there is no 
presence of an explicit
virtual soft-photon, but electromagnetic effects will appear in the
expressions of the scattering lengths as differences
of charged and neutral pseudoscalar masses.  
We proceed the analysis considering the $\pi^- K^+\rightarrow \pi^0 K^0$
transition where
in addition explicit exchange of virtual photons should be considered.

The paper is organized as follows: in sec.~\ref{sec: Leff} we review
briefly the inclusion of electromagnetic corrections
inside the framework of effective lagrangeans, 
emphasizing the role of the low-energy 
constants. We continue with the isospin decomposition for the $\pi-K$
scattering amplitudes,
analyzing the scattering lengths at leading order in the 
isospin limit and comparing them with the isospin breaking corrections in
sec.~\ref{isop}. Next, in sec.~\ref{sec: ONE} 
we proceed with the analysis of the $\pi^0 K^0 \rightarrow \pi^0 K^0$
process at next-to-leading order, reviewing 
first the kinematics and carefully explaining 
how to deal
with isospin breaking effects, strong and electromagnetic ones, in order
to have an  expression  consistent with the chiral power counting.
In sec.~\ref{TWO} we turn to the evaluation of the experimental mode
$\pi^- K^+ \rightarrow \pi^0 K^0$ emphasizing the soft-photon
contribution, the extraction of the Coulomb pole at threshold and the
proper definition of observables once isospin breaking corrections are
taken into account.
In a more technical section, sec.~\ref{sec:TH}, we explain how to 
perform
the threshold expansion of the non-Coulomb part of the scattering amplitude.
In 
sec.~\ref{sec: Results} we review the experimental and theoretical
status of the $\pi-K$ S-wave scattering lengths and we present
our results discussing them
in terms of all input parameters. We make use of our findings to determine
the lifetime of $A_{\pi K}$ in
sec.~\ref{sec: Perstectives}.
Sec.~\ref{sec: Conclusions} summarizes our results. 
Finally, for not interrupting 
the discussion we have collected in the appendices
all relevant 
expressions.

\section{The effective lagrangean to lowest order}
\label{sec: Leff} 
\renewcommand{\theequation}{\arabic{section}.\arabic{equation}}
\setcounter{equation}{0}
This section covers briefly the inclusion of electromagnetic
corrections in a systematic way in the low-energy theory
describing hadron interactions \cite{urech,NR95}. 
Due to the smallness
of the electromagnetic constant, $\alpha_{em}$, these effects have been
theoretically neglected so far in the $\pi K$ scattering process, 
but is well known that near threshold isospin breaking effects
can enhance considerably some observables.  

In presence of electromagnetism it is convenient to split
the lowest order effective lagrangean 
in three terms 
\begin{equation}
\label{lagg}
{\cal L}_2 = {\cal L}^\gamma + {\cal L}_{QCD}^{(2)} + {\cal L}^C\,.
\end{equation}
The foregoing lagrangean possesses the same symmetry restrictions as
the one in the strong sector. Additionally one has to impose an \emph{extra} symmetry,
charge conjugation, affecting only the photon fields and the spurions (see
below). The first term in eq. (\ref{lagg}) corresponds to the usual Maxwell
lagrangean containing the classical 
photon field, $A_\mu$, and the field strength 
tensor, $F_{\mu \nu}$
\begin{equation}
{\cal L}_\gamma = -\frac{1}{4} F_{\mu \nu} F^{\mu \nu} -
\frac{1}{2}\, (\partial_\mu A^\mu)^2\,, \quad F_{\mu \nu} =
\partial_\mu A_\nu - \partial_\nu A_\mu\,.
\end{equation}
The second term formally describes the dynamics
of the strong interaction sector 
\cite{GL2} and 
is given at lowest order by
\begin{equation}
\label{qcd}
{\cal L}_{QCD}^{(2)} 
= \frac{F_0^2}{4} \langle d^\mu U^\dagger d_\mu U + \chi^\dagger
U + \chi U^\dagger \rangle\,.
\end{equation}
As usual brackets, $\langle \ldots \rangle$, stand for trace over
flavour. The field $u(x)$, parametrises the dynamics of the 
low-energy modes in terms of elements of the Cartan subalgebra \cite{callan}. 
The covariant derivative is slightly modified with respect to the
pure QCD interaction expression to accommodate the  
electromagnetic
field
\begin{equation}
d_\mu U = \partial_\mu U - i \left( v_\mu + Q_R A_\mu + a_\mu\right) U + i U
\left( v_\mu+Q_L A_\mu -a_\mu\right)\,,
\end{equation}
$Q_R$ and $Q_L$ are the aforementioned spurions fields, containing the
sources for the electromagnetic operators $A_\mu \overline{q}_L
\left(\frac{\lambda_a}{2}\right) \gamma^\mu q_L$ and $A_\mu \overline{q}_R
\left(\frac{\lambda_a}{2}\right) \gamma^\mu q_R$. Furthermore, from now on
we set them to their constant value
\begin{equation}
Q_L(x) = Q_R(x)=Q, \quad Q = \frac{e}{3} \times \mbox{diag}(2,-1,-1)\,.
\end{equation}
While, as usual, $a_\mu$ and $v_\mu$ stand
for the axial and vector sources respectively. The scalar and pseudoscalar
sources are contained inside the SU(3) matrix $\chi$ as
\begin{equation}
\label{chi}
\chi = s + ip  = 2 B_0 {\cal M} + \ldots\,,
\quad {\cal M} = \mbox{diag}(m_u,m_d,m_s)\,.
\end{equation}
At lowest order the 
last term in eq. (\ref{lagg}), ${\cal L}^C$, determines the masses of the
mesons in the chiral limit, which are of purely electromagnetic
origin. This means that even at tree level the pole of
the two-point Green function is shifted from its QCD value
modifying therefore the kinematics of the low-energy region
\begin{equation}
\label{Cc}
{\cal L}^C = C \langle QUQU^\dagger \rangle = -2 e^2 \frac{C}{F_0^2} \bigl(\pi^-
\pi^+ + K^- K^+ \bigr)+
\cdots\,.
\end{equation}
Obviously the coupling is taken to be universal, thus we expect the same
contribution to the masses  
for charged pions and kaons.
Furthermore, for simplification, the results will be presented in
terms of 
\begin{equation}
Z = \frac{C}{F_0^4} = \frac{M_{\pi^\pm}^2 - M_{\pi^0}^2}{2 e^2 F_0^2}\,.
\end{equation}
The inclusion of isospin breaking
terms can be seen in a very na\"{\i}ve way as coming through two different
sources in eq. (\ref{lagg}). First a pure  \emph{strong} isospin breaking,
i.e. $m_u \ne m_d$. And second an \emph{electromagnetic}
interaction, $e \ne 0$, eq. (\ref{Cc}). In view of the seemingly different role of both
contributions one has to relate them in a consistent way. For instance, the
lagrangean eq. (\ref{lagg}) involves an expansion in several parameters: $p,
m, e$ and $\delta$. Where $p$ refers to the external momenta, $m$ to the
quark-masses, $e$ to the electric charge and finally $\delta = m_d-m_u$. For
being consistent, eq. (\ref{lagg}) should contain operators with the same
chiral order in the series expansion, thus a possible solution can be the
choice $m  \sim e^2 \sim {\cal O}(p^2)$. At the next-to-leading order, 
${\cal O}(p^4)$, none of the following terms are impeded to appear by 
chiral  power counting: $p^4$, $m^2$, $p^2\, m$, $p^2$, $e^2$, $m\, e^2$, 
$p^2\, \delta$, $m\, \delta$, $e^4$ and $\delta^2$, although there is quantitative
support to the assumption, often used in phenomenological discussions, that
the $e^4$ and $\delta^2$ contributions are tiny and can therefore be
safely disregarded in front of the rest. 

Hitherto we have listed all possible electromagnetic, lowest
order operators. Once quantum fluctuations are considered using vertices from
the functional (\ref{lagg}) results are ultraviolet divergent. Those
divergences depend of the regularisation method employed in the loops
diagrams. As is customary we shall adopt the
modified
$\overline{\mbox{MS}}$ subtraction scheme. In order to remove these
ultraviolet divergences, higher order operators, modulated by simple constants,
should be incorporated into the theory with the guidance of the previously
mentioned symmetry requirements. This allows to deal with a theory which is
ultraviolet finite order by order in the parameter expansion and hopefully it
adequately describes several observables.\\
These modulated constants are order parameters of the effective theory 
(low-energy constants) and in the case at hand are determined by the underlying
low-energy dynamics of QCD and QED. For instance at lowest order the order
parameters are given by $F_0$ (eq. (\ref{qcd})), $B_0$ (eq. (\ref{chi})) and
$C$ (eq. (\ref{Cc})). Describing the lowest pseudoscalar decay constant, the
vacuum condensate parameter and the electromagnetic pion mass in the
chiral limit respectively. For the strong  SU(3) sector, at next chiral
order, there are ten new low-energy constants \cite{GL2},
$L_1,\ldots,L_{10}$ and two \emph{high}-energy constants. At this level of
accuracy the ten low-energy constants may be extracted almost independently 
one from each other by matching some observables with the
corresponding experimental determination.
In
the  electromagnetic SU(3) sector there is also need of
higher order operators, up to $16$ modulated via $K_1,\ldots,K_{16}$, to
render any observable free of ultraviolet divergences. 
To gain some information
on these low-energy constants
one has to resort to models \cite{BB97}, to sum-rules
\cite{Bachir} or to a simple crude order estimates.
All in all, one can see that the inclusion of electromagnetic corrections to
hadronic processes increases enormously the number of low-energy constants thus 
washing out any predictability. 

\section{Isospin breaking corrections at leading order}
\renewcommand{\theequation}{\arabic{section}.\arabic{equation}}
\setcounter{equation}{0}
\label{isop}

Under strong interactions symmetry
pions are assigned to a triplet of states,
``\emph{isotriplet} states'',  
whereas kaons can be collected into doublets. This means that the
pion and the kaon are isospin eigenstates with 
eigenvalues $1$ and $1/2$ respectively. Therefore, the amplitudes for the 
$\pi - K$ scattering processes are solely described in terms of two independent 
isospin-eigenstates amplitudes $T^{1/2}$ and $T^{3/2}$  
\begin{eqnarray}
\label{mat}
{\cal M}(\pi^0K^0\rightarrow\pi^0K^0) &
=& \frac{1}{3}\,T^{\frac{1}{2}}\,+\,\frac{2}{3}\,T^{\frac{3}{2}}\,, \nonumber \\
{\cal M}(\pi^-K^+\rightarrow\pi^0K^0) &=
& -\frac{\sqrt{2}}{3}\,T^{\frac{1}{2}}\,+\,\frac{\sqrt{2}}{3}\,T^{\frac{3}{2}}\,, \nonumber \\
{\cal M}(\pi^-K^+\rightarrow\pi^-K^+) &=& \frac{2}{3}\,T^{\frac{1}{2}}\,
+\,\frac{1}{3}\,T^{\frac{3}{2}}\,, \nonumber \\
{\cal M}(\pi^+K^+\rightarrow\pi^+K^+) &=& T^{\frac{3}{2}}\,. 
\end{eqnarray}
It is obvious that under $s \leftrightarrow u$ crossing
the last two matrix elements are
related. In particular one finds
\begin{equation}
\label{rela}
T^{1/2}(s,t,u) = \frac{3}{2} T^{3/2}(u,t,s) - \frac{1}{2} T^{3/2}(s,t,u)\,,
\end{equation}
thus in the isospin limit it is sufficient to compute one of the processes.
It is convenient to use, instead of the invariant amplitudes $T^I$, the partial 
wave amplitudes $t_l^I$ defined in the $s$-channel by
\begin{equation}
\label{partial}
T^I(s,\cos\theta) = 32 \pi \sum_l (2l+1) P_l(\cos\theta) t_l^I(s)\,,
\end{equation}
or by the inverse expression
\begin{equation}
t_l^I(s)\,=\,\frac{1}{64\pi}\,\int_{-1}^{+1}d(\cos\theta)\,P_l(\cos\theta )\,
T^I(s,\cos\theta)\,,
\end{equation}
where $l$ is the total angular momentum, $\theta$ the scattering angle
in the center of mass frame and $P_l$ are the Legendre polynomials with $P_0(\cos\theta )\,=\,1$.
Near threshold the partial wave amplitudes can be parametrized in terms 
of the scattering
lengths, $a_l^I$ and slope parameters, $b_l^I$.
In the normalization (\ref{partial}) the real part of the partial wave amplitude reads  
\begin{equation}
\label{ab}
\textrm{Re}~t_l^I(s) = q^{2l} \left\{a_l^I + b_l^I q^2 + {\cal O}(q^4)
\right\}\,,
\end{equation}
with $q$ being the center of mass three-momentum.

Let us estimate the scattering lengths in the isospin limit at tree level.
Using for instance the first two processes in (\ref{mat}) one can disentangle
the values of each scattering length separately.
In order to match the prescription of the scattering lengths
given in \cite{wein} we define them in terms of the 
charged pion and kaon masses. They read\footnote{We use $F_0 = F_\pi = 93.4$~MeV. See Sec.~(\ref{sec: Results}) 
for the rest of values.}
\begin{equation}
\label{tree}
a_0^{1/2}\,=\,\frac{M_{\pi^{\pm}}M_{K^{\pm}}}{16\pi F_0^2}\,= 0.157~ 
(0.129)\,, \qquad 
a_0^{3/2}\,=\,-\frac{M_{\pi^{\pm}}M_{K^{\pm}}}{32\pi F_0^2}\,= 
-0.079~ (-0.064)\,,
\end{equation}
where the first quoted number refers to the choice $F_0^2 = F_\pi^2$ while the 
second is for $F_0^2 = F_{\pi}F_K$\footnote{In the sequel we shall use
$\frac{F_K}{F_\pi} = 1.22\,.$}. 

When
switching-on the isospin-breaking effects, we are not allowed
anymore to refer to the scattering lengths in a given
isospin-state and new terms in addition to the previous ones arise. 
Furthermore in principle relations as (\ref{rela}) do not hold anymore. This
failure can be seen already at tree level in ${\cal O}(e^2)$ and ${\cal O}(\epsilon)$ terms.
The modified scattering lengths in presence of isospin breaking
can be split as 
\begin{eqnarray}
a_0(00;00)
 &=& \frac{1}{3}\,a_0^{1/2}+\frac{2}{3}\,a_0^{3/2}+\Delta a_0(00;00)\,, \nonumber \\
a_0(+-;00)
&=& -\frac{\sqrt{2}}{3}\,a_0^{1/2}+\frac{\sqrt{2}}{3}\,a_0^{3/2}+\Delta a_0(+-;00)\,, \nonumber \\
a_0(+-;+-)
&=& \frac{2}{3}\,a_0^{1/2}+\frac{1}{3}\,a_0^{3/2}+\Delta a_0(+-;+-)\,, \nonumber \\
a_0(++;++)
&=& a_0^{3/2}+\Delta a_0(++;++)\,, \nonumber  
\end{eqnarray}
where $a_0^{1/2}$ and $a_0^{3/2}$ denote the strong (isospin limit)
S-wave scattering lengths and 
$\Delta a_0(i,j,;l,m)$ represents the leading correction to the corresponding combination 
of scattering lengths due to the isospin breaking effects. 
The evaluation of these 
corrections is straitforward and can be read from the scattering amplitude at leading order
\begin{eqnarray}
\begin{array}{cclcccc}
\Delta a_0(00;00)
  &=& \frac{1}{32\pi F_0^2}\left [
     \frac{1}{4}\,(\Delta_{\pi}-\Delta_K)+
 \frac{\epsilon}{\sqrt{3}}\,(\Mpi +\MK )\right ]&=&0.0032&(0.0026)\,, \nonumber \\
\Delta a_0(+-;00)
  &=& \frac{1}{32\pi\sqrt{2}F_0^2}\bigg [-\frac{\epsilon}{\sqrt{3}}\,(\Mpi +\MK )& & & \nonumber \\
&  &- \frac{\Delta_{\pi}}{4}\,\frac{3 M_K+5M_{\pi}}{M_{\pi}+M_K}+
        \frac{\Delta_K}{4}\,\frac{M_{\pi}-M_K}{M_{\pi}+M_K}\bigg ]&=&-0.0016&(-0.0013)\,,
 \nonumber \\
\Delta a_0(+-;+-)
  &=& \frac{\Delta_{\pi}}{32\pi F_0^2}&=&0.0014&(0.0012)\,, \nonumber \\
\Delta a_0(++;++)
  &=& \frac{\Delta_{\pi}}{32\pi F_0^2}&=&0.0014&(0.0012)\,,          
\end{array}
\end{eqnarray}
with
$
\Delta_m = M^2_{m^\pm}-M^2_{m^0}\,.
$
Notice that whereas in the isospin limit the combination for the scattering lengths cancels in
the $\pi^0 K^0 \rightarrow \pi^0 K^0$ process this is no longer true when isospin breaking
terms are considered. 
In the $\pi^- K^+ \rightarrow \pi^0 K^0$ process the isospin breaking in the 
combination of the scattering lengths is 
roughly a couple of orders
of magnitude smaller than the leading isospin limit quantity.
Even though the future 
experimental bounds on the combination of scattering lengths
for this process is quite restrictive being 
worthwhile to control higher order
corrections.
For the rest of processes the isospin effects are also rather small,
roughly two order of magnitude less than the isospin limit counter parts. 

\section{$\pi^0 K^0 \rightarrow \pi^0 K^0$ process}
\label{sec: ONE}
\renewcommand{\theequation}{\arabic{section}.\arabic{equation}}
\setcounter{equation}{0}
Let us start considering the following process
\begin{equation}
\label{neutral}
\pi^0(p_{\pi_1}) K^0(p_{K_1}) \to \pi^0(p_{\pi_2}) K^0(p_{K_2})\,. 
\end{equation}
Our aim in this section will be to compute its amplitude taking
into account all possible isospin breaking effects. 
Even if the process has not the same experimental interest as the 
$\pi^- K^+ \rightarrow \pi^0 K^0$ reaction it is worth to be 
considered because both processes share almost  the 
same features and complications, exception of the one photon exchange
contribution (see sec.~\ref{SFP}).

\subsection{Kinematics}
\renewcommand{\theequation}{\arabic{section}.\arabic{equation}}
The amplitude for the process, eq. (\ref{neutral}),
can be studied on general grounds in terms of the Mandelstam variables
\begin{equation}
s = (p_{\pi_1}+p_{K_1})^2\,, \quad 
t = (p_{\pi_1}-p_{\pi_2})^2\,, \quad
u = (p_{\pi_1}-p_{K_2})^2\,.
\end{equation}
In the isospin limit and at lowest order of perturbation theory (corresponding 
to the PCAC results), see diagram (a) in fig. 1,
 the off-shell amplitude is given by \cite{wein,Cronin}
\begin{equation}
\label{treeA}
{\cal M} (s,t,u) = \frac{1}{6F^2_0}
\left\{  M_K^2+ M_\pi^2- \frac{u +s}{2} +t 
\right\}\,.
\end{equation}

It is worth to review briefly the kinematics of the process that will be needed
subsequently.
In the center of mass frame the Mandelstam variables are defined in terms
of $q$ and $\theta$
by 
\begin{eqnarray}
s &=& (\,\sqrt{\Mpin \,+\,q^2}\,
+\,\sqrt{M_{K^0}^2\,+\,q^2}\,)^2\,, \nonumber \\
t &=& -2\,q^2\,(\,1\,-\,\cos \theta \,)\,, \nonumber \\
u &=& (\,\sqrt{\Mpin \,+\,q^2}\,
-\,\sqrt{M_{K^0}^2\,+\,q^2}\,)^2\,
-\,2\,q^2\,(\,1\,
+\,\cos \theta \,)\,. \nonumber
\end{eqnarray}

\begin{figure}[t]
\label{diag1}
\leavevmode
\begin{center}
\includegraphics[width=14cm]{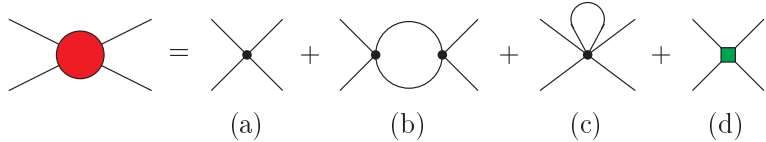}
\end{center}
\caption{
Irreducible topologies.
Vertices in (a), (b) and (c) comes from either eq. (\ref{qcd}) or 
eq. (\ref{Cc}). The vertex in diagram (d) renders the amplitude ultraviolet
finite.}
\end{figure}

\subsection{General framework}
\label{general}
\renewcommand{\theequation}{\arabic{section}.\arabic{equation}}
Hitherto we have considered the process eq. (\ref{neutral}) 
at leading order. In this section we shall sketch the role of isospin
breaking at next-to-leading order.
Indeed, as has been mentioned in the introduction, 
the isospin violating terms that are retained in the 
$\pi-K$ scattering differ from the ones in the $\pi-\pi$ 
reaction due
to the inclusion of the s-quark. 
The later process can be described fully 
in terms of SU(2) quantities where, for instance,
the difference between the charged and neutral  
pion masses are order $\delta^2$ and thus can be disregarded, 
whereas in the former there exist intermediate
strangeness states like $\pi-K$ that give contributions of order $e$ and 
$\delta$. 

To be consistent with the chiral  power counting one has to take into
account all possible scenarios. For instance, given a generic $2 \to 2$
reaction mediated via
the diagram (b) in 
fig.~1 there are two different possibilities of
incorporating isospin violating terms\,: (i) consider
that one of the vertices breaks isospin through the $e^2$ terms 
or via the quark-mass difference, thus at the order we shall work the other 
vertex and the two propagators are taken in the isospin limit, 
or (ii) if both vertices are taken in the isospin limit that forces to 
consider the splitting between charged and neutral
masses (in a given channel) in the same triplet for the pions 
or in the doublet for the kaons in the propagators. Thus within this
prescription we shall consider in the chiral series 
terms up to including $\delta$ and $e^2$
corrections besides the usual $p^4$ at next-to-leading order.

As a matter of fact, the previous distinction, disentangling \emph{strong} and
\emph{electromagnetic} contributions to the isospin breaking terms, is quite
artificial as one can realize when the pseudoscalar masses are
rewritten in terms of bare quantities. Even though it constitutes
a great conceptual help because ultraviolet divergences involving $e^2$ and
$\delta$ terms do not mix at this order allowing to keep track of each
term independently.

For the case we are interested in, involving only neutral particles, there is 
no direct contribution from virtual photon loops, therefore the amplitude is 
safe of infrared singularities and all $e^2$ dependence is due to 
the e.m. mass difference of mesons or by the integration 
of hard photon loops. Hence to obtain the amplitude
including all ${\cal O}(\alpha_{em})$ corrections one needs to restrict
the evaluation to the one-particle-irreducible
diagrams depicted in fig.~1 corresponding  
to the Born amplitude, (a), unitary contribution, (b), tadpole, (c) and finally
the counter-term piece (d). For the precise expressions of this last 
contribution we refer the reader to the
original literature \cite{GL2,urech}. 

To ascertain the correctness 
of our expression we look at the scale independence of the 
result once all  
contributions of the one-particle-irreducible diagrams
are added, the wave function renormalization for the 
external field are taken into account and the $\pi-\eta$ mixing is treated
correctly (see below). Furthermore, when restricting the expressions to 
the isospin limit we recover
the results given in \cite{BKM1}.

Once the amplitude is finite in terms of bare quantities we have to renormalize
the coupling constant, $F_0$, and the masses appearing at lowest
order. 
For the latter contribution
we obtain agreement with the results quoted in \cite{GL2} for the
terms up to including $\epsilon$ corrections and with \cite{urech} for the
electromagnetic ones. While
for the former we shall use two choices:
the first one is to fully renormalize $F_0^2$ as
$F_{\pi}^2$ and for comparison purposes
as 
the combination $F_{\pi}F_K$. 
To this end
we use the
isospin limit quantities \cite{GL2}
\beqa
\label{fpi}
F_{\pi} = F_0 \left\{ 1 + 4
\frac{M_\pi^2+2 M_K^2}{F_0^2}L_4^r +4 \frac{M_\pi^2}{F_0^2} L_5^r  
-2 \mu_\pi -
\frac{1}{2 } \mu_K \right\}\,,
\eeqa
and
\beqa
\label{fpifk}
F_{K} = F_0 \left\{ 1 + 
\frac{M_\pi^2+2 M_K^2}{F_0^2}L_4^r +4 \frac{M_\pi^2}{F_0^2} L_5^r  
-\frac{3}{4} \mu_\pi -
\frac{3}{2} \mu_K -\frac{3}{4} \mu_\eta \right\}\,.
\eeqa
Being $\mu_P$ the finite part of the well-known tadpole integral. 
We refrain to use 
$F_{\pi^0}$ in the numerical estimates of $F_{\pi}$ because experimentally 
is quite poorly  known and instead we shall make use 
of the charged decay constant value. 

The latest contribution enters through
the $\pi-\eta$  mixing. At lowest order 
the mixing angle is given by
\begin{equation}
\label{epsilon}
\epsilon = 
\epsilon^{(2)} \equiv \frac{\sqrt{3}}{4} \; \frac{m_d - m_u}{m_s - \hat{ m}}\,.
\end{equation}
Notice that given the order of accuracy we are considering, $\epsilon^{(2)}$
does not suffice and the next-to-leading order contribution,
$\epsilon^{(4)}$, to
the mixing
angle needs to be considered. We shall use the same approach as 
in \cite{masses} where we refer for
a more detailed explanation. It consists essentially in diagonalyzing
the mixing matrix at the lowest order redefining in that way
the $\pi^0$ and $\eta$ fields. While 
higher order terms in the mixing are 
treated by direct computation of the
S-matrix off-diagonal elements. 
This procedure
is equivalent to the one outlined in \cite{toni}.

Taking 
into account all mentioned contributions  we obtain the renormalized S-matrix
element, ${\cal M}^{00;00}$,  
for the transition $\pi^0 K^0 \rightarrow \pi^0 K^0$ that is gathered in 
app. 
\ref{appdxpoko} where we refer the reader for a detailed exposition. 

\section{$\pi^- K^+ \rightarrow \pi^0 K^0$ process}
\label{TWO}
\renewcommand{\theequation}{\arabic{section}.\arabic{equation}}
\setcounter{equation}{0}
In this section we shall consider the 
more relevant process
\begin{equation}
\label{cc}
\pi^-(p_{\pi^-}) K^+(p_{K^+}) \to \pi^0(p_{\pi^0}) K^0(p_{K^0})\,, 
\end{equation}
with the following Mandelstam variables
\begin{equation}
s = (p_{\pi^-}+p_{K^+})^2\,, \quad 
t = (p_{\pi^-}-p_{\pi^0})^2\,, \quad
u = (p_{\pi^-}-p_{K^0})^2\,. 
\end{equation}
In the center of mass frame these variables read 
\begin{eqnarray}
s   &=& (\,E_1\,+\,E_2\,)^2\,, \nonumber \\
t   &=& \Mpic \,
+\,\Mpin \,-\,2\,E_1\,\sqrt{\Mpin \,+\,{q^\prime}^2}\,+\,2\,q\,q^\prime\,\cos \theta\,, \nonumber \\
u   &=& \Mpic \,+\,M_{K^{0}}^2\,-\,2\,E_1\,\sqrt{M_{K^{0}}^2\,+\,{q^\prime}^2}\,
-\,2\,q\,q^\prime\,\cos \theta\,, 
\end{eqnarray}
with
\begin{eqnarray}
E_1^2 &=& {\Mpic \,+\,q^2}\,, \nonumber \\
E_2^2 &=& {M_{K^\pm}^2\,+\,q^2}\,, \nonumber \\
{q^\prime}^2  &=& {\displaystyle\frac{1}{4}\left [E_1\,
+\,E_2\,+\,\displaystyle\frac{\Delta_{\pi^0K^0}}{\Delta_{\pi^\pm K^\pm}}\,(
          \,E_1\,-\,E_2\,)\,\right ]^2\,-\,\Mpin}\,, \label{eq:c.m. mand. var. for c to n}
\end{eqnarray}
and being $q (q^\prime)$  the
three-momentum of the charged (neutral) particles.  

As has been pointed out earlier, the relevance of this process is
intimately related to the lifetime of the $\pi-K$ system.
Even though we want to stress that our formalism only allows us to
deal with free, on-shell external
particles, in clear contrast with the $\pi-K$ atom where the states are
bounded and off-shell \cite{harup,bernisses}. 
We shall not pursue here the more complete
approach. 

{F}or the construction of most of the graphs (those equivalent
to fig.~1) we shall use the same arguments presented in the 
preceding section.
For the remaining ones 
we shall sketch their treatment in the next section.

\subsection{Soft photon contribution}
\renewcommand{\theequation}{\arabic{section}.\arabic{equation}}
\label{SFP}

\begin{figure}[t]
\leavevmode
\begin{center}
\includegraphics[width=14cm]{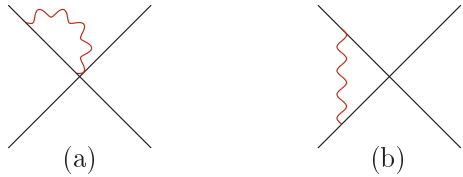}
\end{center}
\hspace{0.5cm}
\caption{
Soft photon contributions to the process $\pi^- K^+ \rightarrow \pi^0 K^0$.
Diagram (a) has a crossed-term. The photon interchanged in diagram (b) 
is only possible between the initial states.} \label{fig3}
\end{figure}

In the case of the $\pi^- K^+ \rightarrow \pi^0 K^0$ process, besides the corrections 
due to the mass difference of the up and down quarks and those generated by the integration
of hard photons 
one has to consider 
corrections due to virtual photons. At order 
$e^2p^2$ these 
corrections arise from the wave function 
renormalization of the charged particles just as from 
the one-photon exchange diagrams depicted in
fig.~2.  
The result of 
diagram (a) in this figure reduces at threshold  
to combinations of
polynomials and logarithms. Whereas the second, (b), needs a closer consideration.
It develops at threshold (see below) a singular behaviour. 
This singularity is issued from the ultraviolet 
finite three-point 
function $C$ defined by 
\begin{equation} 
\label{Co}
C(M_P^2,M_Q^2,m_{\gamma}^2;p_1,p_2)
\equiv  \frac{1}{i}\int\frac{d^dl}{(2\pi )^d}\,\frac{1}
          {(l^2-m_{\gamma }^2)[(p_1-l)^2-M_P^2][(p_2-l)^2-M_Q^2]}\,,
\end{equation}
with the on-shell conditions
$
p_1^2\,=\,M_P^2\,, p_2^2\,=\,M_Q^2 
$
and $m_\gamma^2$ acting as an infrared cut-off for the photon mass. 
Using standard techniques, the integral (\ref{Co}) can be expressed in terms of logarithms 
and dilogarithms. 
{F}or $p^2\,\equiv \,(p_1-p_2)^2 > (M_P+M_Q)^2$, 
and $m_\gamma^2 \rightarrow 0$ it is given as 
follows
\begin{eqnarray}
&&
32\pi^2\lambda _{PQ}^{\frac{1}{2}}(p^2)C(M_P^2,M_Q^2,m_{\gamma}^2;p_1,p_2)
 =  \nonumber \\ &&
-\bigg\{\log\bigg [\frac{\lambda _{PQ}(p^2)}{p^2m_{\gamma }^2}\bigg ] 
- \frac{1}{2}\log\bigg [
         \frac{[\Delta _{PQ}-\lambda _{PQ}^{\frac{1}{2}}
         (p^2)]^2-p^4}{[\Delta _{PQ}+\lambda _{PQ}^{\frac{1}{2}}
         (p^2)]^2-p^4}\bigg ]-i\pi\bigg\}\bigg\{\log\bigg [\frac{[p^2-\lambda _{PQ}^{\frac{1}{2}}
         (p^2)]^2-\Delta _{PQ}^2}{[p^2+\lambda _{PQ}^{\frac{1}{2}}
         (p^2)]^2-\Delta _{PQ}^2}\bigg ]+2i\pi\bigg\} \nonumber \\ 
&&+ 2\textrm{Li}_2~\bigg [\frac{p^2+\Delta _{PQ}+\lambda _{PQ}^{\frac{1}
      {2}}(p^2)}{p^2+\Delta _{PQ}-\lambda _{PQ}^{\frac{1}{2}}
      (p^2)}\bigg ]-2\textrm{Li}_2~\bigg [
      \frac{p^2-\Delta _{PQ}-\lambda _{PQ}^{\frac{1}{2}}
      (p^2)}{p^2-\Delta _{PQ}+\lambda _{PQ}^{\frac{1}{2}}(p^2)}\bigg ]\,, 
      \label{eq:three point function expression}
\end{eqnarray}
where
$$
\Delta_{PQ}\equiv M_P^2-M_Q^2\,, \quad \lambda_{PQ}(p^2)\equiv [p^2-(M_P+M_Q)^2][p^2-(M_P-M_Q)^2]\,,
$$
and finally the dilogarithm function is defined as
$$
\textrm{Li}_2~(z)\equiv -\int_0^zdt\,\frac{\log (1-t)}{t}\,.
$$
The contribution of the $C$ function via the diagram (b) in fig.~\ref{fig3} to the amplitude is
[c.f. eq.~(\ref{eq:one photon exchange contribution})]
$$
\frac{e^2}{\sqrt{2}F_0^2}\,(s-u)(s-\Sigma_{\pi K})C(\Mpi ,\MK ,m_{\gamma}^2;p_{\pi^-},-p_{K^+})\,.
$$
Expanding the real part of the preceding function in the vicinity of the threshold 
by the use of eqs.~(\ref{eq:three point function expression}) and 
(\ref{eq:c.m. mand. var. for c to n}) 
one obtains a Coulomb type behaviour, i.e. $q^{-1}$. Then schematically the threshold 
expansion of the real part of the amplitude takes 
the following form 
\begin{equation} \label{eq:pole substraction}
\textrm{Re}~{\cal M}^{-+ ;00}(s,t,u)\,=\,-\frac{M_{\pi^\pm}M_{K^\pm}}{\sqrt{2}F_0^2}\,\frac{e^2}{4}\,\frac{\mu_{\pi K}}{q}
+\textrm{Re}~{\cal M}_{\textrm{thr.}}^{-+ ;00}+{\cal O}(q)\,,
\end{equation}
with
\begin{equation}
\label{reduced}
\mu_{\pi K}\,=\,\frac{M_{\pi^\pm}M_{K^\pm}}{M_{\pi^\pm}+M_{K^\pm}}\,,
\end{equation}
being the reduced mass of the $\pi-K$ system. 

A remarkable feature of the threshold expansion of the scattering amplitude 
is that neither $\textrm{Re}~{\cal M}_{\textrm{thr.}}^{-+ ;00}$ nor the long-range force
of the photon exchange is affected by the infrared singularity, which only
contributes to ${\cal O}(q^2)$ or to higher orders terms.
In fact, the scattering amplitude contains 
two infrared divergent contributions. The first one comes from the wave 
function renormalization of the charged particles and can be 
read from the Born-type amplitude (\ref{eq:born a for charged}). The second infrared divergent 
piece is due to diagram (b) in fig.~\ref{fig3} and is contained in the 
$C$ function (\ref{eq:three point function expression}). 
Adding both contributions one has the following infrared
piece for the real part of the 
amplitude
\begin{eqnarray}
\textrm{Re}~{\cal M}^{-+ ;00}_{\textrm{ir}}
= \left (\frac{s-u}{\sqrt{2}F_0^2}\right )\left [
\frac{e^2}{16\pi^2}\,\log (m_{\gamma}^2) \right ] 
\left\{1+\frac{1}{2}\,\frac{s-\Sigma_{\pi K}}
{\lambda_{\pi K}^{\frac{1}{2}}(s)}\,
\log\left(
\frac{\left [s-\lambda_{\pi K}^{\frac{1}{2}}(s)\right ]^2-\Delta_{\pi K}^2}
{\left [s+\lambda_{\pi K}^{\frac{1}{2}}(s)\right ]^2
-\Delta_{\pi K}^2}\right)\right\} \,. \nonumber
\end{eqnarray}
As is expected
when evaluated at threshold, this expression vanishes rendering 
$\textrm{Re}~{\cal M}_{\textrm{thr.}}^{-+ ;00}$ as an infrared finite quantity. 
Although the scattering lengths defined in this way will be infrared finite
the slope parameters will not. 
In order to define an infrared finite observable 
one should notice that the cancellation of infrared divergences
takes place order by order in $\alpha_{em}$. This requires
that besides
the virtual 
photon corrections, to take into account real soft photon emission
from the external particles.
Notice that the experimental result will include this bremsstrahlung effect.
In our case is just sufficient to consider one single photon emission, which
amplitude reads
\begin{eqnarray}
&&\hspace{-0.7cm} {\cal M}^{-+;00\gamma}= \frac{e}{2 \sqrt{2} F_0^2}  \epsilon_\mu(k_\gamma)
\left[
-2 (p_{\pi^0}-p_{K^+})^\mu  \right.
\nonumber \\ &&\hspace{-0.7cm} 
+
(p_{K^+}+p_{K^0})\cdot(p_{\pi^+}+p_{\pi^0}-k_\gamma) \,
\frac{(2 p_{\pi^+} - k_\gamma)^\mu}{m_\gamma^2 -2  p_{\pi^+} \cdot  k_\gamma}
- (p_{\pi^+}+p_{\pi^0})\cdot(p_{K^-}+p_{K^0}-k_\gamma)\, 
\frac{(2 p_{K^-} - k_\gamma)^\mu}{m_\gamma^2 -2  p_{K^-} \cdot  k_\gamma} \left.
\right]\,, \nonumber
\end{eqnarray}
with $k_\gamma$ and $\epsilon_\mu(k_\gamma)$ being the momenta of the photon and its polarization 
vector respectively.
As a result, one can write the infrared finite cross-section including 
all ${\cal O}(\alpha_{em})$ but
neglecting ${\cal O}(\alpha^2_{em})$ terms as
\begin{equation} \label{eq:cross section}
\sigma(s;\Delta E) = \sigma^{-+;00}(s)+\sigma^{-+;00\gamma}(s;\Delta E)\,,
\end{equation}
where $\Delta E$ corresponds to the detector resolution.
Once this is done, the corrected scattering 
length $\tilde{a}_0(+-;00)$ might be defined from the threshold 
expansion of the infrared finite 
cross-section, eq.~(\ref{eq:cross section}), by subtracting the Coulomb 
pole term and excluding the corrections due to the mass 
squared differences in the phase-space in the following way \cite{RS76}
\begin{equation} \label{eq:scattering length}
\sigma(s;\Delta E)\,=\,\frac{1}{32\pi s}\,\frac{\lambda_{\pi^0K^0}^{\frac{1}{2}}(s)}
{\lambda_{\pi^-K^+}^{\frac{1}{2}}(s)}\left\{-\,\frac{M_{\pi^{\pm}}M_{K^{\pm}}}{\sqrt{2}F_0^2}
\,\frac{e^2}{4}\,\frac{\mu_{\pi K}}{q}+32\pi \tilde{a}_0(+-;00)+{\cal O}(q)\right\}^2\,.
\end{equation}
We have checked
that the corrections due to the real, soft photon
emission are negligible and thus we expect that the corrected scattering length $\tilde{a}_0(+-;00)$   
will only differ beyond our accuracy from the one obtained using 
eq.~(\ref{eq:pole substraction}), that is, from 
the infrared finite real part of the scattering amplitude at 
threshold.  

Finally, in order to
present our results we shall collect in a single,
infrared
 finite expression
(denoted by $ {\cal M}_{\textrm{ Soft photon}}$ in the tables)
the contributions at threshold of eqs.~(\ref{tadc}, \ref{tadd}, 
\ref{eq:one photon exchange contribution}).

\section{Threshold expansion}
\label{sec:TH} 
\renewcommand{\theequation}{\arabic{section}.\arabic{equation}}
\setcounter{equation}{0}

Let us explain how we obtain the scattering lengths from the scale 
invariant amplitudes eqs.~(\ref{AMPneutral}) and (\ref{AMPcharged}). 
Since we are only interested in the S-wave threshold parameters it is 
just sufficient to expand 
the scattering amplitude
around the threshold values. Even though, this step is not quite 
straightforward because 
in order
to match the prescription given in \cite{wein} for the scattering lengths
we need to shift isospin limit and 
neutral masses to the corresponding charged ones. 
The procedure is rather cumbersome and we shall
use the following substitutions for the masses
\begin{equation}
\label{dell}
M_{\pi^0}^2 \rightarrow M_{\pi^\pm}^2 - \Delta_\pi\,,\quad
M_{K^0}^2 \rightarrow M_{K^\pm}^2 - \Delta_K\,,
\end{equation}
where $\Delta_i$ are small quantities that in the case of pions 
contain at leading order $e^2$ pieces while for kaons contain 
both $e^2$ and $\epsilon$
terms. It is therefore sufficient to expand all quantities
up to first
order in $\Delta_i$.  
For instance in the charged $\rightarrow$ neutral
transition we obtain for the kinematical variables 
\begin{eqnarray}
\label{kin}
&& s= (M_{\pi^\pm} +M_{K^\pm})^2\,, \nonumber \\
&& t = 
-\frac{M_{K^\pm} }{M_{\pi^\pm} + M_{K^\pm} } \Delta_\pi
-\frac{M_{\pi^\pm} }{M_{\pi^\pm} + M_{K^\pm} } \Delta_K\,, \nonumber \\
&& u = (M_{\pi^\pm} -M_{K^\pm})^2
-\frac{M_{\pi^\pm} }{M_{\pi^\pm} + M_{K^\pm} } \Delta_\pi
-\frac{M_{K^\pm} }{M_{\pi^\pm} + M_{K^\pm} } \Delta_K\,. 
\end{eqnarray}
Once this step is performed  and in 
order to book-keep properly the power counting  
any charged mass multiplying $\Delta_i$ or $\epsilon$ pieces
is settled to its isospin limit  
\begin{equation}
\label{kk}
\frac{M_{\pi^\pm}}{M_{\pi^\pm}+M_{K^\pm}} \approx \frac{M_\pi}{M_\pi+M_K} \left(1+
{\cal O}(\epsilon)+ {\cal O}(e^2) \right)\,.
\end{equation}
Only for estimating higher order corrections we shall eventually
keep charged masses in the ratios introduced in eq.~(\ref{kin}). 

Even if the outlined procedure for shifting the neutral masses is 
rather involved 
it allows to expand all one loop 
integrals in an analytic form with quite compact expressions. 
For instance in the s-channel for the neutral $
\rightarrow$ neutral transition we obtain after performing the mentioned steps
the following expansion
\begin{eqnarray}
&&\hspace{-0.8cm}{\mbox {Re} }\bar{J}(\Mpin,\MKn;(M_{\pi^0}+M_{K^0})^2) =
\frac{1}{16 \pi^2} \left[ 1+ \frac{M_{\pi^\pm} M_{K^\pm}}{\MKc -\Mpic}
\log\left(\frac{\MKc}{\Mpic}\right) \right] \nonumber \\ && \hspace{-0.8cm}
-\frac{1}{32 \pi^2} \frac{1}{(M_K^2 -M_\pi^2)^2} 
\left(\frac{M_K}{M_\pi}\Delta_\pi-\frac{M_\pi}{M_K} \Delta_K\right) 
\left[-2(M_K^2 -M_\pi^2)+(M_\pi^2 +M_K^2) \log\left(\frac{M_K^2}{M_\pi^2}\right)\right]\,.\nonumber 
\end{eqnarray}
{F}or loop functions involving the 
$\eta$ mass the use of 
Gell-Mann--Okubo relation
reduces its expression considerably.

In the neutral $\rightarrow$ neutral process,
although the kinematics allows  $t\propto q^2 \approx 0$ at threshold, 
there is no need to consider the 
expansion of the $\bar{J}$ function
in powers of $q$.
This is the case because
all channels contributions behave like $ \mbox{polynomial} \times \bar{J}$  
without any inverse power of kinematical variables ($t$ in this case). 
This does not turn out to be the case in the charged $\rightarrow$
neutral
transition.  
There, one deals with terms 
like $\overline{J}/t$, where in the isospin
limit $t \propto q^2 \rightarrow 0$.  
Expanding near threshold 
$\bar{J}[m_1^2,m_2^2;t(q^2)] \approx  b q^2 + \ldots$ 
we shall obtain contributions from the terms linear in $q^2$. 
Besides this last 
remark there is no more differences in the treatment of the two processes.   

Due to the relevance of the process $\pi^- K^+ \rightarrow \pi^0 K^0$
we
collect, besides the expression of the scattering amplitude, app.~\ref{XX}
the expression of the S-wave scattering lengths in app.~\ref{scl}.

\section{Results and discussion}
\label{sec: Results}
\renewcommand{\theequation}{\arabic{section}.\arabic{equation}}
\setcounter{equation}{0}

Let us first point out that due to the hight threshold of production,
$\sqrt{s_{th}} \sim 632$~MeV, it is not necessary true that a single one-loop
calculation is enough to approach the physical values for the 
scattering lengths. Even though and due to
the fuzzy existing $\pi-K$ data we consider that this can only be
answered once the size of the next-to-next--to-leading order is computed.
Also there are opening of intermediate particle productions, for instance $K \bar K$ in the 
t-channel, that presumably affects strongly the chiral series convergence.

\subsection{Input parameters}
\renewcommand{\theequation}{\arabic{section}.\arabic{equation}}
\setcounter{equation}{0}
Before presenting our results we want to stress the relevance of the
$\pi-K$ scattering process. Its importance goes beyond the determination
of some threshold quantities, but is a touchstone in the knowledge of
spontaneous symmetry breaking. 
Up to nowadays chiral perturbation theory has been used to \emph{parametrize}
the low-energy QCD phenomenology. 
Lacking of enough processes to determine
all low-energy constants one has to resort in theoretical inputs (or prejudices). For instance
the determination of the low-energy constants in the 
electromagnetic sector have in general a quite mild impact on the results and
therefore have been relegated to a secondary place and only recently 
received some attention \cite{BB97,Bachir} due to the increasing precision in the
experiment. But very little is known about them
exception of model estimates. In our treatment 
these constants can play an important role, and
we include them as given in \cite{BB97}, 
where they are estimated by means of resonance saturation. 
(Hereafter all our results are given at the scale $\mu=
M_\rho$).
\begin{eqnarray}
\begin{array}{llll}
K_1^r = -6.4 \cdot 10^{-3}\,,\quad& K_2^r = -3.1 \cdot 10^{-3}\,,\quad&
K_3^r = 6.4 \cdot 10^{-3}\,,\quad &K_4^r = -6.2 \cdot 10^{-3}\,,\nonumber\\
 K_5^r = 19.9 \cdot 10^{-3}\,,\quad& K_6^r = 8.6 \cdot 10^{-3}\,,\quad&
K_7^r\ldots K_{10}^r = 0\,,\quad &K_{11}^r = 0.6 \cdot 10^{-3}\,,\nonumber\\
K_{12}^r = -9.2 \cdot 10^{-3}\,,\quad& K_{13}^r = 14.2 \cdot 10^{-3}\,,
&K_{14}^r = 2.4 \cdot 10^{-3}\,.
\end{array}
\end{eqnarray}
If  instead we use a na\"ive
dimensional analysis the value assigned to each of them 
 would have been restricted to be inside the range
\[
\vert K_i^r \vert\, \lapprox \frac{1}{16\pi^2}\,,
\]
which is taken as a crude indication on the error.
Notice that the central values quoted in 
\cite{BP,Bachir} lie inside this error band.

Contrary to the previous case the low-energy constants in
the strong sector are better known. In a series 
of works \cite{GL2,BG} most of the next-to-leading low-energy constants were pinned down. In addition
to the experimental data a large-N$_c$ arguments were used to settle the marginal
relevance of some operators (those entering together with $L_4$ and $L_6$).
The use of $\pi-K$ data in the $T^{3/2}$ channel can 
disentangle (in principle) the value of
$L_4^r$, due to its product with $M_K^2$ which enhances its sensitivity
to the role of $m_s$ 
\cite{ABB}.

In order to have a more complete control over our results
we use two different set of constants \cite{masses}.
The first one was obtained by fitting simultaneously
the next-to-leading expressions of 
the meson masses, decay constants and the threshold values of the
$K_{\ell 4}$ form-factors to their experimental values\footnote{
Notice that in the decay $K \rightarrow \pi \pi \ell \nu$
the s quark is involved. Thus although if in principle it can be used to obtain 
the value of $L_4^r$ the form-factors $F$ and $G$ turn out
to be rather insensitive to its
actual value
\cite{masses}.}.We shall refer
to it as {\bf set I} and is given by\footnote{Quantities with a star are theoretical inputs.}
\begin{eqnarray}
\begin{array}{lll}
10^3\cdot L^r_1=            0.46\pm 0.23\,,\quad&
10^3\cdot L^r_2=            1.49\pm 0.23\,,\quad&
10^3\cdot L^r_3=         -3.18\pm 0.85\,,\quad  \nonumber \\
10^3\cdot L^r_4=         0\pm 0.5^{*}
\,,\quad& 
10^3\cdot L^r_5=            1.46\pm 0.2\,,\quad&
10^3\cdot L^r_6=            0\pm 0.3^{*}\,,\quad \nonumber \\
10^3\cdot L^r_7=         -0.49 \pm 0.15\,,\quad&
10^3\cdot L^r_8=            1.00\pm 0.20\,. 
\end{array}
\end{eqnarray}
The second set ({\bf set II}) is obtained with the same inputs and
under the same assumptions
as the previous one, but this time the fitted expressions are  
next-to-next-to-leading quantities
\begin{eqnarray}
\begin{array}{lll}
10^3\cdot L^r_1=            0.53\pm 0.25\,,\quad&
10^3\cdot L^r_2=            0.71\pm 0.27\,,\quad&
10^3\cdot L^r_3=         -2.72\pm 1.12\,,\quad  \nonumber \\
10^3\cdot L^r_4=         0\pm 0.5^{*}\,,\quad& 
10^3\cdot L^r_5=            0.91\pm 0.15\,,\quad&
10^3\cdot L^r_6=            0\pm 0.3^{*}\,,\quad \nonumber \\
10^3\cdot L^r_7=         -0.32 \pm 0.15\,,\quad&
10^3\cdot L^r_8=            0.62\pm 0.20\,, 
\end{array}
\end{eqnarray}
As one can see comparing both sets the central value of some of the low-energy 
constants are sizeable shifted from one to another.  
We stress at this point 
that the error in the determination of the scattering lengths is 
mainly associated with the errors 
on low-energy constants. The other quantities
involved in the calculation are hadron masses and for those there are rather
accurate determinations. 
For the latter we use \cite{PDG}
\begin{eqnarray}
\begin{array}{lll}
M_{\pi^\pm} = 139.570~\mbox{MeV}\,,&
M_{\pi^0} = 134.976~\mbox{MeV}\,, \nonumber\\
M_{K^\pm} = 493.677~\mbox{MeV}\,,&
M_{K^0} = 497.672~\mbox{MeV}\,. 
\end{array}
\end{eqnarray}
Notice that in principle 
there is no need of considering $M_\eta^2$ as an additional input. 
It always appears through loop propagators and therefore is sufficient to
consider it 
via the
Gell-Mann--Okubo relation in the presence of isospin breaking
\begin{equation}
\label{GMO}
\Delta_{GMO}\equiv
M_\eta^2 + M_{\pi^0}^2 - \frac{2}{3}\left(M_{K^\pm}^2+M_{K^0}^2\right)
-\frac{2}{3}M_{\pi^\pm}^2 = 0\,.
\end{equation}
Even though we shall also use the value
$M_\eta = 547.30~\mbox{MeV}$ and consider the difference as an indication
of higher order corrections.

Furthermore we make 
use of the isospin limit quantities $M_\pi$ and $M_K$ that can be
defined through combinations of the physical masses
\[
M_\pi^2 = M_{\pi^0}^2\,,\quad M_K^2 = \frac{1}{2}(\MKc+\MKn+ \gamma [\Mpin-\Mpic] 
)\,.
\]
The factor $\gamma$ will take into account any deviation from Dashen's theorem 
\cite{Dashen}. At lowest order ($\gamma =1$) the e.m. relation between the
pseudoscalar masses reads
\[
\left(\MKc-\MKn\right)\Bigm|_{\rm \mbox{\small{e.m.}}} = \Mpic-\Mpin\,.
\]
Next-to-leading contributions to the previous relation
can be quite sizeable. 
As an indication we shall  use 
\cite{BP}
\begin{equation}
\label{dash}
\left(\MKc-\MKn\right)\Bigm|_{\rm \mbox{\small{e.m.}}} = (1.84 \pm 0.25)\left(
\Mpic-\Mpin\right)\,.
\end{equation}
The only remaining input is $F_\pi$ and we shall discuss its value below.  

\subsection{Scattering lengths}
\renewcommand{\theequation}{\arabic{section}.\arabic{equation}}
The studies on the $\pi-K$ scattering
started quite earlier \cite{wein,Cronin} within a current algebra
approach and followed latter by a series of works based on dispersion relations by means
of unitarity and crossing-symmetry \cite{disper1}
(see also \cite{disper1}-\cite{disper4} for recent works using
the same technique).
With the advent of chiral perturbation 
theory the process was analyzed once more \cite{BKM1}. 
Nowadays its interest has been revival
and new approaches 
like the inverse method amplitude \cite{ima}-\cite{ima3} or like the treatment
of kaons
as heavy particles \cite{roessl} have been considered. {\it Grosso modo} 
all mentioned techniques lead to a fairly unique prediction for the scattering
lengths, 
inside the range $[0.16,0.24]$ for 
$a_0^{1/2}$ and $[-0.05,-0.07]$ for
$a_0^{3/2}$. The corrections to the current algebra values 
[c.f. eq.~(\ref{tree})]
are roughly $20\%$ for $a_0^{1/2}$ and
$30\%$ for $a^{3/2}_0$, thus being in the ball park of the usual shifts
between the next-to-leading and  
leading order quantities in processes where chiral perturbation applies.

Even if it seems that from the theoretical point of view there is some general
{\it consensus} the experimental results on the scattering lengths 
are more spread.
In the earlier experiments
most of the collected data on $\pi-K$ were obtained via the 
scattering of kaons on proton or neutron targets. After the data
were analyzed  
by determining the contribution of the one pion exchange. This technique
does not allow to obtain the initial pion on-mass-shell and hence some 
extrapolation is needed. Even if one can perform this extrapolation
the approach is model dependent. The obtained central
values
are $a_0^{1/2} \in [0.168,0.335]$ \cite{mati}--\cite{esta} and
$a_0^{3/2} \in [-0.072,-0.14]$ \cite{esta}--\cite{jong}.   
Latter experiments analyzed the reaction near the threshold 
using dispersive techniques. Even if the 
experimental results are slightly improved  with
respect
to the oldest ones
the threshold parameters are only known within a factor 2 \cite{lang}--\cite{kara}.
A more interesting analysis was performed in \cite{kara}. There,
using the forward sum-rule and measured phase shifts
a direct determination of  
the combination $\vert a^\frac{1}{2}-a^\frac{3}{2} \vert$ 
was obtained  with the result
\begin{equation}
\label{kara}
0.21 \le a^{1/2}-a^{3/2} \le 0.32\,.
\end{equation}
This will be used as a reference point for us. 

\begin{table}[t]
\label{tabneutral}
\begin{center}
\vspace{0.25cm}
\begin{tabular}{cccccc}
\hline
\vspace{0.05cm}
          &  $ a^{1/2}_0+2 a^{3/2}_0$  & $\epsilon$    & $\Delta_\pi$         & $\Delta_K$           & $e^2$ \\
\hline
Tree      & --                  & $0.0052$             & $0.0010$             & $0.0034$             & -- \\
Born(a)   & $-0.0034\,(0.0138)$ & $-0.0020\,(-0.0025)$ & $-0.0006\,(-0.0007)$ & $-0.0019\,(-0.0011)$ & $0$ \\
Born(b)   & $0.0081\,(0.0113)$  & $0.0019\,(0.0012)$   & $0\,(-0.0012)$       & $-0.0001\,(-0.0002)$ & -- \\
Born(c)   & --                  & --                   & --                   & --                   & $0$ \\
Mixing    & --                  & $0.0012$             & --                   & $0$                  & -- \\
$F_\pi^2$ & --                  & $0.0013\,(0.0014)$   & $0.0002\,(0.0002)$   & $0.0009\,(0.0009)$   & -- \\
Tad-pole  & $-0.0447$           & $0.0020$             & $0$                  & $-0.0007$            & --  \\
s-channel & $0.0506$            & $-0.0008$            & $-0.0015$            & $0.0036$             & -- \\
t-channel & $-0.0001$           & $-0.0008$            & $0$                  & $0$                  & -- \\
u-channel & $0.0575$            & $-0.0013$            & $-0.0018$            & $0$                  & -- \\
\hline
          & $0.0679\,(0.0885)$  & $0.0067\,(0.0056)$   & $-0.0025\,(-0.0038)$ & $0.0052\,(0.0059)$   & $0$ \\
\hline
\end{tabular}
\end{center}
\caption{Different chiral contributions to 
the combination of scattering lengths proportional to $a_0(00;00)$. 
We have chosen to renormalize the coupling constant as $F_\pi^2$ with the 
value $F_\pi=93.4$~MeV. In parentheses we show the values obtained with {\bf set I}.
}
\end{table}

Let us turn now to discuss our findings. We shall begin commenting
briefly on the neutral $\rightarrow$ neutral case. Just for illustrative purposes we use
the value $F_\pi = 93.4$ MeV which contains partially some
effects due to electromagnetic corrections. As we shall see none of
our conclusions is affected by this choice. 
In table 1 we have collected
the disentangled contributions to each of the terms given in 
app.~\ref{appdxpoko} with $F_\pi^2$ as renormalization
choice for the coupling constant. We show the isospin limit contribution 
(first column) and
the different corrections to it. Notice that by the definitions (\ref{dell}) of $\Delta_i$
in terms of physical masses, 
$\Delta_\pi$ is just an $e^2$
piece while $\Delta_K$ contains $Z e^2$ and $\epsilon$ terms. This is
the reason why in most of the cases the $\Delta_K$ contribution 
is enhanced with respect to the $\Delta_\pi$ one.
The first thing to remark is that isospin breaking corrections are roughly one order 
of magnitude smaller than the isospin limit quantities. It is worth
to remind that the isospin limit correction comes purely from
the next-to-leading terms. We have kept the physical eta mass
inside the loop diagrams.
In parentheses are quoted the contributions corresponding to
{\bf set I} instead
of {\bf set II}. 
The final result is given by
\begin{equation}
\label{resn1}
3 a_0(00;00) = 
a^{1/2}_0 + 2 a^{3/2}_0 + 3 \Delta_0(00;00) = 
0.0679 + 0.0094 \pm 0.0511\, (0.0885+0.0077 \pm 0.0462)\,,
\end{equation}
renormalyzing
$F_0^2$ as $F_\pi^2$. The first quoted number corresponds to
the isospin limit and the second to the isospin breaking corrections.
Interesting enough are the assigned error that wash out
any sensitivity with respect to the choice of the set
or by any of the choices in the 
renormalization of the coupling constant (see below). 
They come through the uncertainty in the 
low-energy constants and we have propagate them in quadrature. The dominant contributions come
by far from the low-energy constants $L_3^r$ and $L_4^r$ of the pure strong sector
whereas the electromagnetic ones
give an imperceptible contribution. Unfortunately this is not an experimental mode
because this strong sensitivity would constitute a cross-check in the consistency of
the values for $L_3^r$ and $L_4^r$. For sake of completeness we also show the final result
if instead we renormalize $F_0^2$ as $F_\pi F_K$ 
and using  {\bf set II}
\begin{equation}
\label{resn2}
3 a_0(00;00) = 
a^{1/2}_0 + 2 a^{3/2}_0 + 3 \Delta_0(00;00) = 
0.0456 + 0.0087 \pm 0.0343\,.
\end{equation}
While 
the isospin breaking effects are almost unaffected with respect to eq.~(\ref{resn1})
we want to remark the shift in the central value. 
\begin{table}[t]
\begin{center}
\vspace{0.25cm}
\begin{tabular*}{\textwidth}{*{14}{c@{\hspace{2.5mm}}}c}
\hline
\vspace{0.05cm}
          &  $ a^{1/2}_0-a^{3/2}_0$  & $\epsilon$    & $\Delta_\pi$         & $\Delta_K$           & $e^2$ \\
\hline
Tree      & $0.2408$             & $0.0026$             & $0.0018$             & $-0.0009$            & -- \\
Born(a)   & $-0.1060\,(-0.1389)$ & $-0.0010\,(-0.0014)$ & $-0.0007\,(-0.0008)$ & $-0.0001\,(0.0003)$  & -- \\
Born(b)   & $0.0540\,(0.0867)$   & $0.0006\,(0.0012)$   & $-0.0006\,(-0.0008)$ & $0\,(-0.0005)$       & -- \\
Mixing    & --                   & $0.0010$             & --                   & $0$                  & $0$ \\
$F_\pi^2$ & $0.0664\,(0.0688)$   & $0.0007\,(0.0007)$   & $-0.0008\,(-0.0009)$ & $-0.0003\,(-0.0003)$ & -- \\
Tad-pole  & $0.0415$             & $0.0011$             & $0.0002$             & $-0.0005$            & --  \\
s-channel & $0.0283$             & $0.0013$             & $0.0005$             & $0$                  & -- \\
t-channel & $-0.0312$            & $-0.0004$            & $0.0001$             & $-0.0004$            & -- \\
u-channel & $-0.0265$            & $-0.0009$            & $0$                  & $0.0001$             & -- \\
Soft photon & --                 & --                   & --                   & --                   & $-0.0004$ \\
\hline
          & $0.2674\,(0.2695)$  & $0.0047\,(0.0053)$  & $0.0005\,(0)$      & $-0.0011\,(-0.0020)$ & $-0.0004$ \\
\hline
\end{tabular*}
\end{center}
\caption{\label{tablec1}Different chiral contributions to 
the combination of scattering lengths proportional to $ 
{\mbox Re}\,{\cal M}^{+-;00}$.
We have choose to renormalize $F_0^2$  as $F_\pi^2$ with the 
value $F_\pi=92.4$~MeV. We show the first four digits without any rounding.
Quantities
between brackets correspond to {\bf set I}.
}
\end{table}

Let us turn now to discuss the most interesting mode, the 
charged $\rightarrow$ neutral transition.
The experimental proposal claims an accuracy of $20\%-30\%$ on the 
measurement of  
the lifetime,
this roughly translate in $10\%-15\%$ of 
accuracy for the determination scattering lengths. 
As before we have disentangled all contributions and explored all possible
scenarios  allowed by
the input parameters. In our estimates we shall use the more appropriate value
$F_\pi = 92.4$~MeV which does not contain 
electromagnetic corrections \cite{holdstein}.
Even if the most natural choice 
is $F_\pi F_K$ we
shall proceed as in the previous case and also show the results 
using $F_\pi^2$ as an
indication of the sensitivity to this parameter.

Like this case is by far of greater interest we have refined slightly some 
considerations that we shall bear in mind in the remainder:
$(i)$ we have used the Gell-Mann--Okubo relation for the eta mass inside the
loop functions. $(ii)$  Only kept the first term in the r.h.s. of
eq.~(\ref{kk}).
And $(iii)$ 
we assume Dashen's theorem to hold at next-to-leading order. (We shall
discuss below the uncertainties associated with these considerations.)
Similarly to the previous table the numbers quoted in parentheses
refer to {\bf set I}. 

We have collected the results for $F_\pi^2$ in table \ref{tablec1}. 
As one can see the difference between the two sets 
is at most $2\%$ in the final result. Also the isospin breaking effects 
are roughly two order of magnitude smaller than the isospin limit quantity. 
In this
case it seems that the experimental setup sensitivity is not enough to
detect the new effect we have incorporated.
Adding all partial contributions we obtain
\begin{eqnarray}
\label{resc2}
&& \hspace{-0.8cm}- \frac{3}{\sqrt{2}}a_0(+-;00) = 
a^{1/2}_0 - a^{3/2}_0 - \frac{3}{\sqrt{2}} \Delta_0(+-;00) = 
0.2674 + 0.0037 \pm 0.0022 \,(0.2695 + 0.0033 \pm 0.0023)\,, \nonumber \\&&
\end{eqnarray}
stressing that the contents of the previous expression
is only indicative because the way 
we have chosen to renormalize the coupling constant. Notice that 
the assigned errors are at the same footing
than the isospin breaking 
terms. A closer look to the errors reveals that they have mainly
an electromagnetic
origin. Furthermore they are dominated
clearly by 
$K_{10}^r $ and $K_{11}^r$ low-energy constants. 
In the strong sector the dominant errors come  (by order
of dominant contribution) from $L_5^r$ and $L_8^r$. Those low-energy
constants are strongly correlated.

\begin{table}[t]
\begin{center}
\vspace{0.25cm}
\begin{tabular*}{\textwidth}{*{14}{c@{\hspace{2.5mm}}}c}
\hline
\vspace{0.05cm}
  & $ a^{1/2}_0-a^{3/2}_0 $& $\epsilon$ & $\Delta_\pi$ & $\Delta_K$ & $e^2$ \\
\hline
Tree      & $0.1974$             & $0.0021$             & $0.0015$             & $-0.0008$            & -- \\
Born(a)   & $-0.0712\,(-0.0933)$ & $-0.0007\,(-0.0009)$ & $-0.0004\,(-0.0005)$ & $0\,(0.0002)$        & -- \\
Born(b)   & $0.0363\,(0.0582)$   & $0.0004\,(0.0008)$   & $-0.0004\,(-0.0005)$ & $0\,(0)$             & -- \\
Mixing    & --                   & $0.0007$             & --                   & $0$                  & $0$ \\
$F_\pi\,F_K$ & $0.0705\,(0.0815)$   & $0.0007\,(0.0008)$   & $0\,(0)$             & $-0.0001\,(-0.0001)$ & -- \\
Tad-pole  & $0.0279$             & $0.0007$             & $0.0001$             & $-0.0003$            & --  \\
s-channel & $0.0192$             & $0.0009$             & $0.0003$             & $0$                  & -- \\
t-channel & $-0.0209$            & $-0.0003$            & $0.0001$             & $0.0002$             & -- \\
u-channel & $-0.0179$            & $-0.0006$            & $0$                  & $0.0001$             & -- \\
Soft photon & --                & --                   & --                   & --                   & $-0.0003$ \\
\hline
       & $0.2412\,(0.2520)$   & $0.0038\,(0.0043)$   & $0.0011\,(0.0009)$   & $-0.0009\,(-0.0011)$ & $-0.0003$ \\
\hline
\end{tabular*}
\end{center}
\caption{\label{tablec2}Different chiral contributions to 
the combination of scattering lengths proportional to $ 
{\mbox Re}\,{\cal M}^{+-;00}$.
The entries in the table differs from table~\ref{tablec1} just in the
renormalization of the coupling constant. Here we have used the 
combination $F_\pi F_K$ with the constrain $\frac{F_K}{F_\pi}=1.22$
and the value $F_\pi=92.4$~MeV.
We show the first four digits without any rounding. Quantities
between brackets correspond to {\bf set I}.
}
\end{table}

Our \emph{main} results are collected in table \ref{tablec2}. It 
corresponds to the choice of renormalization $F_\pi F_K$ for the
decay constant.
Once more isospin breaking effects turn out to be two order of magnitude
smaller than the isospin limit quantities. Furthermore, as in table \ref{tablec1} but in a more
accentuated way,
table \ref{tablec2} shows a strong cancellation between the isospin breaking contributions
(see for instance the contribution of $\Delta_\pi$ and $\Delta_K$).
Adding all contributions from the table
one obtains
\begin{equation}
\label{RES}
-\frac{3}{\sqrt{2}}a_0(+-;00) = 
0.2412 + 0.0037\pm 0.0034\, (0.2520 + 0.0038 \pm 0.0043)\,.
\end{equation}
Notice that once more the theoretical errors are small but still 
competitive in size with  
the isospin breaking effects. It also seems that 
with this choice of renormalization of coupling constants
we weight more the role of the low-energy constant
$L_5^r$, (this is reflected in the sizeable uncertainty if we use
{\bf set I}).
It is worth to stress at this point that the expression 
of the amplitude for the charged $\rightarrow$ neutral transition does not
depend on the low-energy constant $L_4^r$ whereas $L_6^r$ only comes 
inside isospin breaking terms. Due to the numerical irrelevance of those last
terms and 
even lacking a complete knowledge on the role of large-N$_c$ suppressed
operators,
our estimates (in that sense) are precise and unambiguous.
The difference between the central values of the two sets are at most $5\%$. 
Hence, without taking into account the error bars, and in
the most
optimistic case, $10\%$ of accuracy in the experimental result,
any sensitivity to the set of low-energy
constants is just borderline.
Before concluding let us comment about the role of using the physical eta
mass, the full expression in the l.h.s in eq.~(\ref{kk}) and  
Dashen's theorem, eq.~(\ref{dash}). 
If instead of using the Gell-Mann--Okubo relation (\ref{GMO}) one uses the physical eta mass, 
the central value in (\ref{RES}) increases by $\sim 0.0003$. Higher order terms as introduced by
eq.~(\ref{kk}) also increases (\ref{RES}) roughly  by an amount $\sim 0.003$. And finally
the shift allowed by
the violation of Dashen's theorem goes beyond the accuracy we quote. A consistent way of incorporating these 
effects 
in our estimates is to consider them 
as a crude guess of higher order corrections and we shall treat them as theoretical uncertainties, therefore we
add all three in quadrature and to the previous error bar in eq.~(\ref{RES}). This leads to the
final estimate for the combination of scattering lengths
\beqa
\label{RE}
- \frac{3}{\sqrt{2}} a_0(+-;00) =
a^{1/2}_0 - a^{3/2}_0 - \frac{3}{\sqrt{2}} \Delta_0(+-;00) 
= \left\{ \begin{array}{l}
{\bf (Set I)}\quad 0.2520+0.0038\pm 0.0073\,.\vspace{0.05in} \\
{\bf (Set II)}\quad 0.2412 +0.0037  \pm 0.0045\,.   \vspace{0.05in}  
\end{array}
\right.
\eeqa

\section{Perspectives}
\label{sec: Perstectives}
\renewcommand{\theequation}{\arabic{section}.\arabic{equation}}
\setcounter{equation}{0}

Using the previous estimates for the combination of scattering lengths, eq.~(\ref{RE}), we can
calculate partially the lifetime of  the $A_{\pi K}$ atoms. 
Inside a relativistic framework the probability of the transition 
of an A$_{\pi K}$ atom 
\[
A_{\pi K} \rightarrow \pi^0 + K^0
\]
can be casted as
\[
W_{n,0}(\pi^0 K^0) \propto 
\left( \textrm{Re}~{\cal M}^{\pm ;0}_{\textrm{thr.}} \right)^2\,.
\]
Focusing on the \emph{isospin limit} the previous relation
is given by \cite{transp}
\begin{equation}
\label{experi}
W_{n,0}(\pi^0 K^0) \approx
\frac{8\pi}{9} \left( \frac{2\Delta m}{\mu_{\pi K}}\right)^{1/2} 
\frac{(a_0^{1/2}-a_0^{3/2})^2 \vert
\Psi_{n,0}(0)\vert^2}{1+\frac{2}{9} \mu_{\pi K} \Delta m 
(a_0^{1/2}+2 a_0^{3/2})^2}
 + \cdots \approx
\frac{1}{\tau_{n,0}}\,,
\end{equation}
where $n$ is the principal quantum number,  
\begin{equation}
\Delta m = (M_{K^\pm}+M_{\pi^\pm})-(M_{\pi^0}+M_{K^0})\,
\end{equation}
and $\mu_{\pi K}$ is given in eq.~(\ref{reduced}).
Finally $\Psi_{n,0}(0)$ is the A$_{\pi K}$ 
Coulomb wave function at the origin and it is given by
\begin{equation}
\vert \Psi_{n,0}(0) \vert^2 = 
\frac{p_B^3}{\pi n^3}\,,\quad p_B=\frac{e^2}{4\pi} \mu_{\pi K}\,.
\end{equation}
Furthermore the orbital angular momentum, $l$, can be safely taken equal
to zero. If one allows $l> 0$ the transition is suppressed by chiral power 
counting, i.e.
turns to be of ${\cal O}(e^4)$.
Notice that in the previous relation (\ref{experi}) enter precisely
both combinations of scattering lengths we have found,
$a^{1/2}_0-a^{3/2}_0$ and $a^{1/2}_0+2 a^{3/2}_0$. Even though
the term added to unity in the denominator of (\ref{experi}) can be neglected,
since it is of order $10^{-5}$, thus beyond the accuracy of our calculation or
the experimental sensitivity and hence only
the combination $a^{1/2}_0-a^{3/2}_0$ 
turns to be
of relevance. This simplify slightly the expression to
\begin{equation}
\label{experi2}
\tau_{1,0} \approx 
\frac{0.274 \cdot 10^{-15}}{M_{\pi^\pm}^2}
\left(\frac{2}{9}\right) a_0(+-;00)^{-2} \,.
\end{equation} 
Inserting eq.~(\ref{RE}) in
eq.~(\ref{experi2}) one finds the following value for the $A_{\pi K}$ 
lifetime  in the ground state
\begin{equation}
\label{tau}
\tau_{1,0} = 4.58 \cdot 10^{-15}~ \mbox{s} \quad 
( 4.20 \cdot 10^{-15}~ \mbox{s})\,,
\end{equation}
where the quantity quoted between brackets corresponds to {\bf set I}. 

Let us stress once more that in
writing (\ref{experi}) 
we have not taken into account any source of isospin breaking and
consequently a theoretical determination of the $\pi-K$ lifetime is just
halfway. Hence the content of eq.~(\ref{tau}) is just \emph{indicative}. 
In fact eq.~(\ref{experi}) is not 
suitable to handle bound state systems being by far the framework
of a non-relativistic lagrangean \cite{lepage} the
most efficient way of treating bound states. 
Even though
without any further control over the errors on the low-energy constants it 
seems
not worth to pursue this analysis.

\section{Summary }
\label{sec: Conclusions}
\renewcommand{\theequation}{\arabic{section}.\arabic{equation}}
\setcounter{equation}{0}

In this work we have estimated the role of isospin breaking effects in the transitions
$\pi^0 K^0 \rightarrow \pi^0 K^0$ and $\pi^- K^+ \rightarrow \pi^0 K^0$. They turn out 
to be rather mild.
Furthermore the former reaction is quite interesting because its sensitivity to the large-N$_c$ suppressed
operator involving $L_4^r$. From a more practical point of view (essentially because the existence of 
experimental data) we have carefully evaluated the shift in the scattering lengths for the 
$\pi^- K^+ \rightarrow \pi^0 K^0$ reaction. While the error from the low-energy constants turns out to be
compatible with the future experimental sensitivity, estimates of higher order corrections  
are at the same footing. Contrary to the previous case this reaction does not contain sizeable
contribution from large-N$_c$ suppressed operators. Bearing in mind 
the results in eq.~(\ref{RE}) and the expected experimental sensitivity
we can conclude that in principle isospin breaking effects
do not affect the determination of the lifetime.
An accuracy in the determination of the $\pi-K$ S-wave scattering
lengths 
at the same level as in pionium
experiments 
will disentangle between the two sets
of low-energy constants
and thus will constitute a major step in understanding the basic structure
of the effective lagrangean.  
As a direct application we have evaluated (partially) the expected lifetime
of the $A_{\pi K}$ atom.

\vskip.4cm
\noindent {\bf Acknowledgements}\\
We thank Ll. Ametller for collaborating in the earlier stages of this project,
discussions and a critical reading of the manuscript. We also express our gratitude to M. Knecht and H. Sadzjian
for discussions and encouragement in pursuing the presented work. 
P.\ T. was supported by EC--Contract No.
ERBFMRX--CT980169.

\vskip.4cm
\noindent {\it Note added }\\
When completing this work, ref.~\cite{KM} appears. 
It contains partially our work. As has been shown the isospin breaking
effects are quite mild and hidden by sizeable error bars. This reference
essentially leads to the same numerical conclusions. Even though after some
partial checks we notice that 
our expressions contain
some log dependence from the three
point function that are missing in the
mentioned reference. 

\appendix
\setcounter{equation}{0}
\newcounter{zahler}
\addtocounter{zahler}{1}
\renewcommand{\thesection}{\Alph{zahler}}
\renewcommand{\theequation}{\Alph{zahler}.\arabic{equation}}
\label{appdx}

\setcounter{section}{0}
\setcounter{subsection}{0}

\renewcommand{\thesection}{\Alph{zahler}}
\renewcommand{\theequation}{\Alph{zahler}.\arabic{equation}}

\setcounter{equation}{0}
\setcounter{zahler}{0}
\addtocounter{zahler}{1}
\renewcommand{\thesection}{\Alph{zahler}}
\renewcommand{\theequation}{\Alph{zahler}.\arabic{equation}}

\section{$\pi^0 K^0 \rightarrow \pi^0 K^0$ scattering amplitude}
\label{appdxpoko}

In this appendix we collect all relevant formulae for the
neutral$\rightarrow$ neutral amplitude.
We are aware that the ``digestion" of this kind of expressions is always
hard, thus 
for sake of clarity we have not mixed (in a good extent) the contributions.
This has probably enlarged slightly the expressions but we find this worth
for any future
comparison.

The amplitude is written as
\begin{eqnarray}
\label{AMPneutral}
{\cal M}(s,t,u) =&& 
\left\{ 
{\cal M}\Bigm|_{\textrm{tree}}	+
	\sum_{i=a,b,c} {\cal M}_{(i)}\Bigm|_{\textrm{born}} 
+
{\cal M}\Bigm|_{\textrm{mixing}}  
+
{\cal M}\Bigm|_{\textrm{tadpole}} 
+
{\cal M}\Bigm|_{\textrm{t-channel}}
\right\} 
\nonumber \\
&&
+
\left\{
{\cal M}\Bigm|_{\textrm{s-channel}}
+
s \leftrightarrow u
\right\}\,, \no \\
\end{eqnarray}
making explicit use of the $s\leftrightarrow u $ symmetry.
In the sequel we discuss briefly the relevant pieces.
Notice also that we explicitly use $F_0$ in at all orders. This stays for the 
\emph{non-renormalized} decay constant.

The leading order contribution is given by
\begin{eqnarray}
{\cal M}\Bigm|_{\textrm{tree}} &=&
    \frac{1}{4F_0^2}(t+\Delta _{\pi}-\Delta _K)+\frac{1}{2F_0^2}
     \bigg (\frac{\epsilon}{\sqrt{3}}\bigg )(s+u-2t)\,. 
\end{eqnarray}

The Born-type term containing the w.f.r. and bare masses
renormalization
contributions is given by
\begin{eqnarray}
{\cal M}_{(a)}\Bigm|_{\textrm{born}} &=&
 \frac{1}{4F_0^2}(t+\Delta _{\pi}-\Delta _K)\bigg\{\,\frac{1}{6}\,(\mupin +
     3\mueta +6\muKn +10\mupic +4\muKc ) \nonumber \\
&-& \frac{8}{F_0^2}\bigg [\,2(\Mpi +2\MK )L_4^r+\Sigma_{\pi K}L_5^r\bigg ] 
     \nonumber \\
&+& \frac{\epsilon}{\sqrt{3}}(\mueta -\mupi )-\frac{16}{F_0^2}
\bigg(\frac{\epsilon}{\sqrt{3}}\bigg) \Delta_{K\pi}L_5^r\,\bigg\}\nonumber  \\
&+& \frac{\MKn}{6F_0^2}\bigg\{-\frac{2}{3}\mueta -
      \frac{8\epsilon}{\sqrt{3}}(\mueta -\mupi )-\frac{16}{F_0^2}\bigg(
\frac{\epsilon}{\sqrt{3}}\bigg)\Delta_{K\pi}(2L_8^r-L_5^r) \nonumber \\
&-& \frac{8}{F_0^2}\bigg [\,(\Mpi +2\MK )(2L_6^r-L_4^r)+\MK (2L_8^r-L_5^r)\,\bigg ]\,\bigg\} 
     \nonumber \\
&+& \frac{\Mpin}{6F_0^2}\bigg\{\,\mupin +\frac{1}{3}\mueta -2\mupic  \nonumber \\
&-& \frac{8}{F_0^2}\bigg [\,(\Mpi +2\MK )(2L_6^r-L_4^r)+\Mpi (2L_8^r-L_5^r)\,\bigg ]\,\bigg\} 
     \nonumber \\
&+& \frac{\Delta _K}{4F_0^2}\bigg\{\frac{2}{3}\mueta +
     \frac{8}{F_0^2}\bigg[
     (\Mpi +2\MK )(2L_6^r-L_4^r)+2\MK (2L_8^r-L_5^r)\,\bigg ]\,
     \bigg\} \nonumber \\
&+& \frac{\Delta _{\pi}}{4F_0^2}\bigg\{\,\frac{\Mpi}{16\pi ^2F_0^2}+
     4\mupi -2\muK -\frac{2}{3}\mueta  \nonumber \\
&-& \frac{8}{F_0^2}\bigg [\,(\Mpi +2\MK )(2L_6^r-L_4^r)+2\MK 
      (2L_8^r-L_5^r)-\Delta _{\pi K}L_5^r\,\bigg ]\,\bigg\}
      \nonumber \\
&+& \frac{1}{2F_0^2}\bigg (\frac{\epsilon}{\sqrt{3}}\bigg )(s+u-2t)
     \bigg\{\,\frac{1}{6}(11\mupi +3\mueta +10\muK ) \nonumber \\
&-&
\frac{8}{F_0^2}\bigg [\,2(\Mpi +2\MK )
L_4^r+\Sigma_{\pi K}L_5^r\,\bigg ]\,\bigg\} \nonumber \\
  &-& \frac{e^2\,t}{18F_0^2}\,\bigg[
        24(K_1^r+K_2^r)-18K_3^r+9K_4^r+14(K_5^r+K_6^r) \bigg] \nonumber \\
  &+& \frac{e^2\MK}{6F_0^2}\,\bigg [\frac{3}{8\pi ^2}-9F_0^2\muKtild  \nonumber \\
  &+& \frac{2}{9}[12( K_1^r+K_2^r-K_7-K_8^r)
        -5K_5^r-5K_6^r-4K_9^r+50K_{10}^r+54K_{11}^r]\,\bigg ] \nonumber \\
  &+& \frac{e^2\Mpi}{6F_0^2}\,\bigg [-\frac{3}{8\pi ^2}+9F_0^2\mupitild \nonumber \\
  &+& \frac{1}{9}[24( K_1^r+K_2^r-K_7-K_8^r)+18K_3^r-9K_4^r+20(K_5^r+K_6^r)-2K_9^r \nonumber \\
  &-& 110K_{10}^r-108K_{11}^r]\,\bigg ]\,,
\end{eqnarray}
where as is customary
\begin{eqnarray}
\quad \Sigma_m = M^2_{m^\pm}+M^2_{m^0}\,,&& \Sigma_{mn}= M^2_{m}+M^2_{n}\,.
\end{eqnarray}

The effect of the $\pi-\eta$ mixing is taken into account by
\begin{eqnarray}
&{\cal M}&\Bigm|_{\textrm{mixing}}  
= \frac{1}{144F_0^2}\,\left [2(s+u-2t)- \Delta_{\pi^0\eta}
  \right ]\bigg\{ -3(\muKc -\muKn ) \nonumber \\
  &-& \frac{36\epsilon}{\sqrt{3}}\,\left (\mupi -\mueta \right )
  -\frac{24\epsilon}{\sqrt{3}}\,\left (\mupi -\muK \right )
+\frac{864\epsilon}{\sqrt{3}F_0^2}\,\Delta_{\pi \eta}
\left (3L_7+L_8^r\right )\bigg\} \nonumber \\
&+& \frac{1}{144F_0^2}\,\left [2\left (\frac{s+u-2t}{\Mpin -\Meta^2}\right )-1\right ]\bigg\{ 12\Sigma_{\pi^0K^0}(\muKc -\muKn ) \nonumber \\
  &-& \frac{96\epsilon}{\sqrt{3}}\,\Mpi\left (\mupi -\muK \right )+8e^2\Mpi \left [3(2K_3^r-K_4^r)-2(K_5^r+K_6^r)+2(K_9^r+K_{10}^r)
        \right ]\bigg\}\,. \nonumber\\
\end{eqnarray}

The Born-type contribution containing one insertion of the 
strong ${\cal O}(p^4)$ counter-terms is written as
\bea
{\cal M}_{(b)}\Bigm|_{\textrm{born}} 
=  \frac{1}{F_0^4}\sum_{i=1}^8{\cal P}_iL_i^r\,,
\eea
with
\begin{eqnarray}
       {\cal P}_1 & = & 8(2\Mpin -t)(2\MKn -t)\,, \nonumber \\
       {\cal P}_2 & = & 4\bigg [\,(\Sigma _{\pi^0K^0}-s)^2
                         +(\Sigma _{\pi^0K^0}-u)^2\,\bigg ]\,, \nonumber \\     
       {\cal P}_3 & = & 2(1-\frac{2\epsilon}{\sqrt{3}})
                         (2\Mpin -t)(2\MKn -t) 
                   +  (1+\frac{2\epsilon}{\sqrt{3}})
                         \bigg [\,(\Sigma _{\pi^0K^0}-s)^2
                         +(\Sigma _{\pi^0K^0}-u)^2\,\bigg ]\,, \nonumber \\
       {\cal P}_4 & = & -\frac{2}{3}\bigg [\,\Mpi +14\MK 
                         -\frac{6\epsilon}{\sqrt{3}}(5\Mpi -2\MK )
                         \,\bigg ]
                         (2\Mpin -t) \nonumber \\
                  & - & \frac{2}{3}\bigg [\,13\Mpi +2\MK 
                         -\frac{6\epsilon}{\sqrt{3}}(\Mpi +2\MK )
                         \,\bigg ]
                         (2\MKn -t) \nonumber \\
                  & + & \frac{2}{3}\,t\bigg [\,\Mpi +2\MK 
                         -\frac{6\epsilon}{\sqrt{3}}(\Mpi +2\MK )
                         \,\bigg ]\,, \nonumber \\
       {\cal P}_5 & = & -\frac{2}{3}\bigg (\,\Mpi +3\MK 
                         -\frac{12\epsilon}{\sqrt{3}}\,\Mpi
                         \,\bigg )
                         (2\Mpin -t) \nonumber \\
                  & - & \frac{2}{3}\bigg [\,3\Mpi +\MK 
                         -\frac{4\epsilon}{\sqrt{3}}(5\Mpi -2\MK )
                         \,\bigg ]
                         (2\MKn -t) \nonumber \\
                  & + & \frac{2}{3}\,t\bigg [\,-\Sigma_{\pi K} 
                         +\frac{4\epsilon}{\sqrt{3}}(2\Mpi -5\MK )
                         \,\bigg ]\,,   \nonumber \\
       {\cal P}_6 & = & \frac{8}{3}\bigg [\,2M_K^4+15\Mpi\MK +M_{\pi}^4
                         +\frac{16\epsilon}{\sqrt{3}}
                         (M_K^4+\Mpi\MK -2M_{\pi}^4)
                         \,\bigg ]\,,
                         \nonumber \\
       {\cal P}_7 & = & \frac{64\epsilon}{\sqrt{3}}
                         (2M_K^4-\Mpi\MK -M_{\pi}^4)\,, \nonumber \\
       {\cal P}_8 & = & \frac{8}{3}\bigg [\,M_K^4+6\Mpi\MK +M_{\pi}^4
                         +\frac{2\epsilon}{\sqrt{3}}
                         (19M_K^4-6\Mpi\MK -13M_{\pi}^4)
                         \,\bigg ]\,.
                         \nonumber 
\end{eqnarray}
The same kind of diagram but with the e.m. counter-term is given by
\begin{eqnarray}
{\cal M}_{(c)}\Bigm|_{\textrm{born}} 
= &-&\frac{2e^2}{27F_0^2}(3K_1^r+3K_2^r+K_5^r+K_6^r)
    (s+u-2t) \nonumber \\
&+& \frac{4e^2}{27F_0^2}
    (3K_7+3K_8^r+K_9^r+K_{10}^r)\,\Sigma _{\pi K}\,. 
\end{eqnarray}
If mesons were really massless, any tadpole contribution type
will vanish, as SU(3) symmetry is broken precisely by
quark masses this does not turns out to be the case in nature. One thus obtains 
\begin{eqnarray}
&{\cal M}&\Bigm|_{\textrm{tadpole}} 
                   =  -\frac{\mupin}{18F_0^2}\bigg [\,t+2\Mpin 
                         -\frac{12\epsilon}{\sqrt{3}}(t+\Mpi -2\MK )\,\bigg ] 
                         \nonumber \\
                  & - & \frac{\mueta}{18F_0^2}\bigg [\,3t-4\MKn 
-\frac{12\epsilon}{\sqrt{3}}(t-2\MK )\,\bigg ] \nonumber \\
                  & - & \frac{\mupic}{9F_0^2}\bigg [\,2(t-3\Mpin )
                         -\frac{\epsilon}{\sqrt{3}}(9t-2\Mpi -4\MK )\,\bigg ] 
                         \nonumber \\
                  & - & \frac{\muKn}{9F_0^2}\bigg [\,3t-2\Mpin 
                         -\frac{6\epsilon}{\sqrt{3}}(3t-2\Mpi -4\MK )
                         \,\bigg ] 
                         \nonumber \\ 
                  & - & \frac{\muKc}{18F_0^2}\bigg [\,t+2\Mpin -8\MKn 
                         -\frac{4\epsilon}{\sqrt{3}}(3t+4\Mpi -4\MK )
                         \,\bigg ]\,. 
\end{eqnarray}
Hitherto we have shown the terms that are polynomials. When performing
the one loop corrections one obtains an unitary piece. This is given in terms
of non-analytical functions, i.e. essentially functions with a
momenta dependence in the argument. In order to present them we have kept
track
on the internal propagators, 
thus the identification of each diagram is straight 
forward. 
They are given in the t-channel by
\[
{\cal M}_{\textrm{t-channel}}=
{\cal M}_{\pi^0\pi^0}+
{\cal M}_{\eta\eta}+
{\cal M}_{\pi^0\eta}+
{\cal M}_{\pi^+\pi^-}+
{\cal M}_{K^0 \bar{K}^0}+
{\cal M}_{K^+ K^-}\,,
\]
where 
\begin{eqnarray}
{\cal M}_{\pi^0\pi^0}
&    =& \frac{\Mpin}{24F_0^4}\bigg\{\,4F_0^2(1-
         \frac{6\epsilon}{\sqrt{3}})\mupin \nonumber \\
    &+& 3\bigg [\,t+\Delta _{\pi}-
         \Delta _{K} -\frac{2\epsilon}{\sqrt{3}}(3t-2
         \Sigma _{\pi K})\,\bigg ]\overline{B}(\Mpin,\Mpin;t)
	 \,\bigg\}\,.
\end{eqnarray}
\begin{eqnarray}
{\cal M}_{\eta\eta}
    &=& \frac{\Mpin}{72F_0^4}\bigg\{\,12F_0^2(1+
         \frac{2\epsilon}{\sqrt{3}})\mueta \nonumber \\
    &+& \bigg [\,9t-6\Meta^2 -2\Mpin +3(\Delta _{K}-\Delta _{\pi})+
          \frac{6\epsilon}{\sqrt{3}}(3t-2
 \Sigma _{\eta K})\,\bigg ]\overline{B}(\Meta^2,\Meta^2;t)\,\bigg\}\,. \nonumber
 \\
\end{eqnarray}
\begin{eqnarray}
{\cal M}_{\pi^0\eta}
    &=& \frac{1}{3F_0^4}\bigg (\,
          \frac{\epsilon}{\sqrt{3}}\,\bigg )\Delta_{K \pi}\bigg [\,
          2F_0^2\mupi +2F_0^2\mueta +(3t-4\MK )\overline{B}(\Mpi,\Meta^2;t)
          \,\bigg ]\,.
\end{eqnarray}
\begin{eqnarray}
{\cal M}_{\pi^+\pi^-}
    &=& \frac{1}{36F_0^4}\bigg [\,
          4F_0^2\mupic (5t-3\Mpin )+9t(t-\Mpin )\overline{B}(\Mpic,\Mpic;t)
          \,\bigg ]\,.
\end{eqnarray}
\begin{eqnarray}
{\cal M}_{K^0 \bar{K}^0}
    &=& \frac{1}{72F_0^4}\bigg\{\,-4F_0^2\muKn\bigg 
          [\,-5t+3(\Delta _{K}-\Delta _{\pi})+
         \frac{6\epsilon}{\sqrt{3}}(5t-2\Sigma _{\pi K})\,\bigg ] \nonumber \\
    &+& 9t\bigg [\,t+\Delta _{\pi}-\Delta _{K}-
          \frac{2\epsilon}{\sqrt{3}}(3t-2
         \Sigma _{\pi K})\,\bigg ]\overline{B}(\MKn, \MKn;t)\,\bigg\}\,.
\end{eqnarray}
\begin{eqnarray}
{\cal M}_{K^+K^-}
    &=& \frac{1}{144F_0^4}\bigg\{\,4F_0^2\muKc\bigg 
          [\,5t+3(\Delta _{K}-\Delta _{\pi})+
         \frac{6\epsilon}{\sqrt{3}}(5t-2\Sigma _{\pi K})\,\bigg ] \nonumber \\
    &+& 9t\bigg [\,t+\Delta _{K}-\Delta _{\pi}+
          \frac{2\epsilon}{\sqrt{3}}(3t-2
         \Sigma _{\pi K})\,\bigg ]\overline{B}(\MKc,\MKc;t)\,\bigg\}\,.  
\end{eqnarray}
{F}or  
the s(u)-channel the intermediate particles contribution is 
\[
{\cal M}\Bigm|_{\textrm{s-channel}} = 
{\cal M}_{\pi^+\pi^-} + 
{\cal M}_{\pi^0 K^0}+
{\cal M}_{\eta K^0}\,.
\]
The above terms are given by
\begin{eqnarray}
{\cal M}_{\pi^+K^-}&=&
\frac{1}{8F_0^4}\bigg\{\,2F_0^2\muKc \,\bigg [\,3s-3\Mpic -\MKc +4\Delta _{\pi}
        +\frac{2\epsilon}{\sqrt{3}}(3s-3\Mpi -\MK )\,\bigg ] \nonumber \\
   &+& \bigg [\,s-\Sigma_{\pi^\pm K^\pm} +2\Delta _{\pi}+\frac{\epsilon}{\sqrt{3}}
        (s-\Sigma_{\pi K})\,\bigg ]^2\overline{B}(\Mpic,\MKc;s)  \nonumber \\
   &+& 2\bigg [\,s-\Sigma _{\pi^\pm K^\pm}+2\Delta _{\pi}
        +\frac{\epsilon}{\sqrt{3}}(s-\Sigma _{\pi K})\,\bigg ]
\nonumber \\	&& \hspace{0.3cm}\bigg [\,
        s-\Delta _{\pi^0K^0}+\frac{\epsilon}{\sqrt{3}}(s+3\Delta _{\pi K})
        \,\bigg ]\overline{B}_1(\Mpic, \MKc;s) \nonumber \\
   &+& \bigg [\,
        s-\Delta _{\pi^0K^0}+\frac{\epsilon}{\sqrt{3}}(s+3\Delta _{\pi K})
        \,\bigg ]^2\overline{B}_{21}(\Mpic,\MKc;s) \nonumber \\
   &+& 2s\bigg [\,
        2\MKn -t-\frac{2\epsilon}{\sqrt{3}}(2u-t-2\Mpi )\,
        \bigg ]\overline{B}_{22}(\Mpic,\MKc;s)\,\bigg\}\,.
\end{eqnarray}
\begin{eqnarray}
{\cal M}_{\pi^0 K^0}&=&
    \frac{1}{144F_0^4}\bigg\{\,6F_0^2\muKn \,\bigg [\,s-\Mpin
        -3\MKn 
        -\frac{4\epsilon}{\sqrt{3}}(3s-3\Mpi -5\MK )\,\bigg ] \nonumber \\
   &+& \bigg [\,5\Mpin +\MKn -s-\frac{6\epsilon}{\sqrt{3}}
        (5\Mpi -3\MK -s)\,\bigg ]^2\overline{B}(\Mpin,\MKn;s) \nonumber \\
   &-& 2(s+3\Delta _{\pi^0K^0})\bigg [\,5\Mpin +\MKn -s
        +\frac{12\epsilon}{\sqrt{3}}(s-5\Mpi +\MK )\,\bigg ]
        \overline{B}_1(\Mpin,\MKn;s) \nonumber \\
   &+& (1-\frac{12\epsilon}{\sqrt{3}})(s+3\Delta _{\pi^0K^0})^2
        \overline{B}_{21}(\Mpin,\MKn;s)  \nonumber \\
   &+& 2s(1-\frac{12\epsilon}{\sqrt{3}})\bigg [\,
       4(u-t+\Delta _{\pi^0K^0})+2\MKn -t\,
        \bigg ]\overline{B}_{22}(\Mpin,\MKn;s)\,\bigg\}\,. 
\end{eqnarray}
\begin{eqnarray}
{\cal M}_{\eta K^0}&=&
   \frac{1}{432F_0^4}\bigg\{\,18F_0^2\muKn \,\bigg [\,3s+\Mpin
        -5\MKn 
        -\frac{2\epsilon}{\sqrt{3}}(6s+6\Mpi -14\MK )\,\bigg ] \nonumber \\
   &+& \bigg [\,3s-7\Mpin +\MKn -\frac{2\epsilon}{\sqrt{3}}
        (3s+\Mpi -7\MK )\,\bigg ]^2\overline{B}(\Meta^2,\MKn;s) \nonumber \\
   &+& 6(s+3\Delta _{\pi^0K^0})\bigg [\,3s-7\Mpin +\MKn 
        -\frac{12\epsilon}{\sqrt{3}}(s-\Mpi -\MK )\,\bigg ]
        \overline{B}_1(\Meta^2, \MKn;s) \nonumber \\
   &+& 9(1-\frac{4\epsilon}{\sqrt{3}})(s+3\Delta _{\pi^0K^0})^2
        \overline{B}_{21}(\Meta^2,\MKn;s) \nonumber \\
   &+&  \hspace{-0.3cm} 18s(1-\frac{4\epsilon}{\sqrt{3}})\bigg [\,
       4(u-t+\Delta _{\pi^0K^0})+2\MKn -t\,
        \bigg ]\overline{B}_{22}(\Meta^2,\MKn;s)\,\bigg\}\,.
\end{eqnarray}
The $\overline{B}_{ij}$ functions are defined in app. \ref{int}.

\setcounter{section}{0}
\setcounter{subsection}{0}

\renewcommand{\thesection}{\Alph{zahler}}
\renewcommand{\theequation}{\Alph{zahler}.\arabic{equation}}

\setcounter{equation}{0}
\addtocounter{zahler}{1}
\renewcommand{\thesection}{\Alph{zahler}}
\renewcommand{\theequation}{\Alph{zahler}.\arabic{equation}}

\section{$\pi^- K^+ \rightarrow \pi^0 K^0$ }

In this appendix we display the relevant formulae concerning 
the more interesting process. As before our philosophy has been not to mix
too much the terms.
One can verify that the following result result
is scale invariant.

\subsection{Scattering amplitude }
\label{XX}

The amplitude is written as
\begin{eqnarray}
\label{AMPcharged}
\hspace{-1cm}{\cal M}(s,t,u)& =& 
\left\{ {\cal M}\Bigm|_{\textrm{tree}}
+  
\sum_{i=a,b,c,d} {\cal M}_{(i)}\Bigm|_{\textrm{born}} 
+
{\cal M}\Bigm|_{\textrm{mixing}}  
+
{\cal M}\Bigm|_{\textrm{tadpole}}
\right. \nonumber \\ 
&+&
\left.
{\cal M}\Bigm|_{\textrm{s-channel}}
+
{\cal M}\Bigm|_{\textrm{t-channel}}
+
{\cal M}\Bigm|_{\textrm{u-channel}}
+
{\cal M}\Bigm|_{\textrm{one-photon}}
\right\}\,.
\end{eqnarray}
In the above expression the splitting between the terms is essentially the 
same as in the pure neutral case, but in addition we have the last term that 
concerns the explicit photon exchange.

For the tree-level amplitude we have
\begin{equation}
{\cal M}\Bigm|_{\textrm{tree}}\,=\,\frac{u-s}{2\sqrt{2}F_0^2}-\frac{\Delta _{\pi}}{2\sqrt{2}F_0^2}
     +\frac{\Delta _K-\Delta _{\pi}}{4\sqrt{2}F_0^2}-\frac{1}{2\sqrt{2}F_0^2}
     \bigg (\frac{\epsilon}{\sqrt{3}}\bigg )(s+u-2t)\,.
\end{equation}

The Born amplitude is given by
\begin{eqnarray}
{\cal M}_{(a)}\Bigm|_{\textrm{born}} &=&
     \frac{u-s}{2\sqrt{2}F_0^2}\bigg\{\,\frac{1}{6}\,(3\mupin +
     3\mueta +5\muKn +8\mupic +5\muKc ) \nonumber \\
&-& \frac{8}{F_0^2}\bigg [\,2(\Mpi +2\MK )L_4^r+\Sigma_{\pi K}L_5^r\bigg ]\bigg\} 
      +\frac{\MK}{\sqrt{2}F_0^2}\left (\frac{\epsilon}{\sqrt{3}}\right )\,
      (\mueta -\mupi )\nonumber \\
&+& \frac{\Delta _K}{4\sqrt{2}F_0^2}\bigg\{\frac{1}{6}(11\mupi -\mueta +
     10\muK ) \nonumber \\
&-& \frac{8}{F_0^2}[(\Mpi +2\MK )(L_4^r+2L_6^r)-\Delta_{K \pi}L_5^r+4\MK 
     L_8^r]
     \bigg\} \nonumber \\
&-& \frac{\Delta _{\pi}}{4\sqrt{2}F_0^2}\bigg\{\frac{3\Mpi}{16\pi ^2F_0^2}+
     \frac{1}{6}(129\mupi +5\mueta +42\muK ) \nonumber \\
&-& \frac{8}{F_0^2}[(\Mpi +2\MK )(3L_4^r+2L_6^r)+2\MK (L_5^r+2L_8^r)]\bigg\}
      \nonumber \\
&-& \frac{1}{2\sqrt{2}F_0^2}\bigg (\frac{\epsilon}{\sqrt{3}}\bigg )(s+u-2t)
     \bigg\{\frac{1}{6}(11\mupi +3\mueta +10\muK ) \nonumber \\
&-&\frac{8}{F_0^2}[2(\Mpi +2\MK )L_4^r+\Sigma_{\pi K} L_5^r]\bigg\}\,. \label{eq:born a for charged}
\end{eqnarray}

As in the neutral to neutral transition, there is a term 
from the treatment of the $\pi-\eta$ mixing 
\begin{eqnarray}
&&{\cal M}\Bigm|_{\textrm{mixing}}  
= -\frac{1}{144\sqrt{2}F_0^2}\,\left [2(s+u-2t)
-\Delta_{\pi^0\eta}
+2\Delta_{\pi}\right ]\bigg\{ -3(\muKc -\muKn ) \nonumber \\
&-& \frac{36\epsilon}{\sqrt{3}}\,\left (\mupi -\mueta \right )
-\frac{24\epsilon}{\sqrt{3}}\,\left (\mupi -\muK \right )
+\frac{864\epsilon}{\sqrt{3}F_0^2}\,\Delta_{\pi\eta}
\left (3L_7+L_8^r\right )\bigg\} \nonumber \\
&-& \frac{1}{144\sqrt{2}F_0^2}\,
\left [2\left (\frac{s+u-2t+\Delta_{\pi}}{\Mpin -\Meta^2}\right )-1\right ]
\bigg\{ 12\Sigma_{\pi^0K^0}(\muKc -\muKn ) \nonumber \\
&-& \frac{96\epsilon}{\sqrt{3}}\,\Mpi
\left (\mupi -\muK \right )+8e^2\Mpi 
\left [3(2K_3^r-K_4^r)-2(K_5^r+K_6^r)+2(K_9^r+K_{10}^r)
\right ]\bigg\}\,.\nonumber\\
\end{eqnarray}

The strong leading lagrangean at ${\cal O}(p^4)$ contributes as 
\begin{equation}
 {\cal M}_{(b)}\Bigm|_{\textrm{born}} 
          =  \frac{2}{\sqrt{2}F_0^4}\sum_{i=3}^8{\cal P}_iL_i^r\,,  
\end{equation}
being 
\begin{eqnarray}
 {\cal P}_3 & = & \frac{2\epsilon}{\sqrt{3}}(2\Mpi -t)(2\MK -t) \nonumber \\
    & + & (1-\frac{\epsilon}{\sqrt{3}})(\Sigma_{\pi^0 K^+} -u)
                        (\Sigma_{\pi^+ K^0} -u) \nonumber \\
                  & - & (1+\frac{\epsilon}{\sqrt{3}})(\Sigma_{\pi^0 K^0} -s)
                        (\Sigma_{\pi^+ K^-} -s)\,, \nonumber \\
{\cal P}_4 & = & -2(\Mpi +2\MK )(s-u)-\frac{2\epsilon}{\sqrt{3}}
                (\Mpi +2\MK )(2\Sigma_{\pi K}-3t)\,, \nonumber \\
{\cal P}_5 & = & -2\Sigma_{\pi K}(s-u)+\frac{6\epsilon}{\sqrt{3}}
               \Sigma_{\pi K}\,t+\frac{4\epsilon}{\sqrt{3}}
                   (M_K^4-M_{\pi}^4-4\Mpi\MK )\,, \nonumber \\
{\cal P}_6 & = & -\frac{8\epsilon}{\sqrt{3}}(2M_K^4-M_{\pi}^4-\Mpi\MK )\,,
\nonumber \\
{\cal P}_7 & = & -\frac{32\epsilon}{\sqrt{3}}(2M_K^4-M_{\pi}^4-\Mpi\MK )\,,
\nonumber \\
{\cal P}_8 & = & -\frac{8}{3}\bigg (\frac{\epsilon}{\sqrt{3}}\bigg )
                      (17M_K^4-7M_{\pi}^4-10\Mpi\MK )\,. \nonumber 
\end{eqnarray}
The polynomials not displayed explicitly do not contribute to the process. 

The equivalent contribution, but in the e.m. sector is casted as
\begin{eqnarray}
\label{tadc}
{\cal M}_{(c)}\Bigm|_{\textrm{born}} 
&=& \frac{2e^2}{3\sqrt{2}F_0^2}(2K_1^r+2K_2^r-4K_3^r+2K_4^r+K_5^r+4K_6^r)
    (\Sigma_{\pi K}-s) \nonumber \\
&-& \frac{2e^2}{9\sqrt{2}F_0^2}(6K_1^r+6K_2^r-6K_3^r+3K_4^r+4K_5^r+4K_6^r)
    (\Sigma_{\pi K}-u) \nonumber \\
&+& \frac{e^2}{9\sqrt{2}F_0^2}(12K_3^r-6K_4^r-K_5^r-K_6^r)
    (2\Mpi -t)\nonumber  \\        
&+& \frac{e^2}{3\sqrt{2}F_0^2}(K_5^r+K_6^r)
    (2\MK -t) \nonumber \\
&-& \frac{2e^2}{9\sqrt{2}F_0^2}
    \bigg [\,9(\Mpi +2\MK )K_8^r-\Mpi K_9^r+(17\Mpi +18\MK )K_{10}^r
      +18\Sigma_{\pi K} K_{11}^r\,\bigg ]\,. \nonumber\\ 
\end{eqnarray}

The $e^2$ contribution from the wave function renormalization
reads
\begin{eqnarray}
\label{tadd}
{\cal M}_{\textrm{(d)}}\Bigm|_{\textrm{born}}
  &=& \frac{e^2(u-s)}{2\sqrt{2}F_0^2}\,\bigg [-2F_0^2\mupitild -2F_0^2\muKtild -\frac{1}{16\pi ^2}\bigg (2+
        \log \frac{m_{\gamma}^2}{\Mpi}+\log \frac{m_{\gamma}^2}{\MK}\bigg )
        \nonumber \\
  &-& \frac{1}{9}(48K_1^r+48K_2^r-18K_3^r+9K_4^r+34K_5^r+34K_6^r)\bigg ]
        \nonumber    \\
  &-& \frac{e^2\MK}{4\sqrt{2}F_0^2}\,\bigg [\frac{1}{4\pi ^2}-6F_0^2\muKtild -\frac{4}{3}(K_5^r+K_6^r+
       12K_8^r
       -6K_{10}^r-6K_{11}^r)\bigg ] \nonumber \\
  &+& \frac{e^2\Mpi}{4\sqrt{2}F_0^2}\bigg [\frac{3}{4\pi ^2}-18F_0^2\mupitild 
        -\frac{2}{3}(18K_3^r-9K_4^r-12K_8^r+2K_9^r-34K_{10}^r-36K_{11}^r)\bigg ]\,, \nonumber \\
\end{eqnarray}
with
\[
\tilde\mu_P = \frac{1}{32 \pi^2 F^2} \log\left(\frac{M^2_P}{\mu^2}\right)\,.
\]

The tadpole diagram contributes as
\begin{eqnarray}
{\cal M}\Bigm|_{\textrm{tadpole}} 
                  & = & \frac{\mupin}{3\sqrt{2}F_0^2}\bigg [\,s-u+\Delta _{\pi}
                         -\frac{\epsilon}{\sqrt{3}}(3t+3\Mpi -4\MK )\,\bigg ] 
                         \nonumber \\
                  & + & \frac{\mueta}{6\sqrt{2}F_0^2}\bigg [\,s-u+\Delta _{\pi}
-\frac{\epsilon}{\sqrt{3}}(3t+2\Mpi -12\MK )\,\bigg ] \no \\
                  & + & \frac{\mupic}{18\sqrt{2}F_0^2}\bigg [\,9(s-u)+
                        3(\Delta _K+4\Delta _{\pi})
                         -\frac{\epsilon}{\sqrt{3}}(15t+4\Mpi -16\MK )\,\bigg ] 
                         \nonumber \\
                  & + & \frac{\muKn}{18\sqrt{2}F_0^2}\bigg [\,8s+5t-4u-4
                        \Sigma_{\pi}+3\Delta _{\pi}
                         +\frac{\epsilon}{\sqrt{3}}(21s+27u-40\Mpi -20\MK )
                         \,\bigg ] 
                         \nonumber \\ 
                  & + & \frac{\muKc}{18\sqrt{2}F_0^2}\bigg [\,4s-5t-8u+4
                        \Sigma_{\pi}+27\Delta _{\pi}
                         +\frac{\epsilon}{\sqrt{3}}(27s+21u-40\Mpi -20\MK )
                         \,\bigg ]\,. \no \\ 
\end{eqnarray}
Like in the neutral case, we display the unitary contribution 
disentangling each term separately. We remind that the subscripts
refer to the internal particles running in the propagators.
For the s-channel one gets
\[
{\cal M}\Bigm|_{\textrm{s-chanel}} = -\frac{1}{12\sqrt{2}F_0^4} \left\{
{\cal M}_{\pi^-K^+}+\frac{1}{2}
{\cal M}_{\pi^0K^0}+\frac{1}{18}
{\cal M}_{\eta K^0}\right\}\,,
\]
where 
\begin{eqnarray}
&&{\cal M}_{\pi^-K^+}
 = 
2F_0^2(1+\frac{\epsilon}{\sqrt{3}})\bigg [\,2(s-\Mpin )+s-
\Delta _{\pi^-K^+}+6\Delta _{\pi} \label{eq}\bigg] \muKc \nonumber \\   
&+& (s+\Delta _{\pi^-K^+}+6\Delta _{\pi})\bigg [\,
           s-\Sigma _{\pi^0K^0}+\frac{\epsilon}{\sqrt{3}}(s-5\Mpi +3\MK )\,
           \bigg ]  \overline{B}(\Mpic, \MKc;s) \nonumber \\ 
&+& \left\{ (s-3\Delta _{\pi^-K^+})\bigg [\,
  s-\Sigma _{\pi^0K^0}+\frac{\epsilon}{\sqrt{3}}(s-5\Mpi +3\MK )\,
        \bigg ]  \right. \nonumber \\
 &+& \left.(s+\Delta _{\pi^-K^+}+6\Delta _{\pi})\bigg [\,
           s-\Delta _{\pi^0K^0}+\frac{\epsilon}{\sqrt{3}}(s+3\Delta_{\pi K} )\,
         \bigg ] \right\}  \overline{B}_1(\Mpic, \MKc;s)  \nonumber \\
 &+& (s-3\Delta _{\pi^-K^+})\bigg [\,
     s-\Delta _{\pi^0K^0}+\frac{\epsilon}{\sqrt{3}}(s+3\Delta_{\pi K} )\,
     \bigg ] \overline{B}_{21}(\Mpic, \MKc;s) \nonumber  \\
 &-& 2s\bigg [\,\Sigma _{\pi^-K^0}-u+2(t-\Sigma _{K^+K^0})
   +\frac{4\epsilon}{\sqrt{3}}(\Sigma _{\pi K}-t)-
  \frac{5\epsilon}{\sqrt{3}}(\Sigma _{\pi K}-u)\,\bigg ] 
   \overline{B}_{22}(\Mpic, \MKc;s)\,. \nonumber \\
\end{eqnarray}
\begin{eqnarray}
&&{\cal M}_{\pi^0K^0}
      =  \nonumber  \\
&-& 2F_0^2(1+\frac{\epsilon}{\sqrt{3}})\bigg [\,3s
     -3\Mpin -5\MKn -\frac{6\epsilon}{\sqrt{3}}
   (3s-3\Mpi -\MK )\,\bigg ] \muKn \nonumber  \\           
 &+& \bigg [\,s-\Sigma _{\pi^0K^0} 
 + \frac{\epsilon}{\sqrt{3}}
  (s-9\Mpi +7\MK )\,\bigg ] \times  \nonumber \\&& \hspace{0.1cm} \bigg [\,
  5\Mpin +\MKn -s-\frac{6\epsilon}{\sqrt{3}}(5\Mpi -3\MK -s)\,
      \bigg ]  
     \overline{B}(\Mpin, \MKn;s) \nonumber \\ 
&+& \left\{ \bigg [\,s-\Delta _{\pi^-K^+} 
   + \frac{\epsilon}{\sqrt{3}}
  (s+3\Delta_{\pi K} )\,\bigg ] \times
  \right. \nonumber \\ && \hspace{0.1cm} \bigg [\,
   5\Mpin +\MKn -s-\frac{6\epsilon}{\sqrt{3}}(5\Mpi -3\MK -s)\,
   \bigg ] \nonumber \\ 
      &-& \left. (1-\frac{6\epsilon}{\sqrt{3}})(s+3\Delta _{\pi^0K^0} )\bigg [\,
           s-\Sigma _{\pi^0K^0} +\frac{\epsilon}{\sqrt{3}}(s-9\Mpi +7\MK )\,
  \bigg ]  \frac{}{} \right\}  \overline{B}_{1}(\Mpin, \MKn;s) \nonumber\\
 &-& (1-\frac{6\epsilon}{\sqrt{3}})(s+3\Delta _{\pi^0K^0} )\bigg [\,
  s-\Delta _{\pi^-K^+} +\frac{\epsilon}{\sqrt{3}}(s+3\Delta _{\pi K})\,
         \bigg ] \overline{B}_{21}(\Mpin, \MKn;s) \nonumber \\
  &-& 2s(1-\frac{6\epsilon}{\sqrt{3}})
           \bigg [\,2(\Sigma _{\pi^0K^+}-u)
      + t-\Sigma _{K^+K^0}
           +\frac{4\epsilon}{\sqrt{3}}(2\Mpi -t) \nonumber \\ &+&
           \frac{\epsilon}{\sqrt{3}}(2\MK -t)-
           \frac{4\epsilon}{\sqrt{3}}(\Sigma _{\pi K}-u)\,\bigg ] 
	    \overline{B}_{22}(\Mpin, \MKn;s)\,.
\end{eqnarray}
\begin{eqnarray}
{\cal M}_{\eta K^0}
      &=&
   18F_0^2\bigg [\,3s
  +\Mpin -5\MKn -\frac{\epsilon}{\sqrt{3}}
        (15s+13\Mpi -33\MK )\,\bigg ] \muKn \nonumber \\           
  &+& \bigg [\,3s-7\Mpin +\MKn   
      - \frac{\epsilon}{\sqrt{3}}
  (9s-5\Mpi -13\MK )\,\bigg ]\times \nonumber \\&& \hspace{0.1cm}\bigg [\,
  3s-7\Mpin +\MKn -\frac{2\epsilon}{\sqrt{3}}(3s+\Mpi -7\MK )\,
        \bigg ] \overline{B}(\Meta^2,\MKn;s) \nonumber \\
      &+& \left\{
      3(1-\frac{2\epsilon}{\sqrt{3}})(s+3\Delta _{\pi^0K^0} )\bigg [\,
           3s-7\Mpin +\MKn -\frac{\epsilon}{\sqrt{3}}(9s-5\Mpi -13\MK )\,
           \bigg ] \right. \nonumber \\
 &+& 3\bigg [\,3s-7\Mpin +\MKn 
   - \frac{2\epsilon}{\sqrt{3}}
  (3s+\Mpi -7\MK )\,\bigg ]\times \nonumber \\ && \hspace{0.1cm} \left.
  \bigg [\,
 s+3\Delta_{\pi^+ K^-} -\frac{3\epsilon}{\sqrt{3}}(s-\Delta_{\pi K} )\,
           \bigg ]  \right\} \overline{B}_1(\Meta^2,\MKn;s) \nonumber \\
 &+& 9(1-\frac{2\epsilon}{\sqrt{3}})(s+3\Delta _{\pi^0K^0} )\bigg [\,
 s+3\Delta _{\pi^-K^+} -\frac{3\epsilon}{\sqrt{3}}(s-\Delta _{\pi K})\,
       \bigg ]  \overline{B}_{21}(\Meta^2,\MKn;s) \nonumber  \\
 &+& 18s(1-\frac{2\epsilon}{\sqrt{3}})
           \bigg [\,2(u-s-t) 
     + 4(\Sigma _{\pi}-t)+\Sigma _K-t \nonumber  \\
          & +& \frac{3\epsilon}{\sqrt{3}}(2\MK -t)-
           \frac{6\epsilon}{\sqrt{3}}(\Sigma _{\pi K}-u)\,\bigg ] 
	   \overline{B}_{22}(\Meta^2,\MKn;s)\,. 
\end{eqnarray}

In the t-channel 

\begin{equation}
{\cal M}\Bigm|_{\textrm{t-chanel}} = {\cal M}_{\pi^-\pi^0}
+{\cal M}_{\pi^-\eta}+{\cal M}_{K^-{\bar K}^0}\,,
\end{equation}
with
\begin{eqnarray}
{\cal M}_{\pi^-\pi^0}
  &=& -\frac{1}{6\sqrt{2}F_0^4}\bigg\{\,\frac{4\epsilon}{\sqrt{3}}
       (3t-\Mpi -4\MK )F_0^2\mu_\pi \nonumber \\
  &+& 2(3\Mpi -t)\bigg [\,\Delta _{\pi}+\frac{\epsilon}{\sqrt{3}}
       (2\Mpi -t)\,\bigg ] \overline{B}(M_\pi,M_\pi;t) \nonumber \\
  &-& 2\bigg [\,3\Mpi \Delta _{\pi}+\frac{\epsilon}{\sqrt{3}}\,t(2\Mpi -t)
       +\frac{\epsilon}{\sqrt{3}}(3\Mpi -t)(t+4\Delta _{\pi K})\,\bigg ]
       \overline{B}_1(M_\pi,M_\pi;t) \nonumber \\
  &+& 2t\bigg [\,\Delta _{\pi}+\frac{\epsilon}{\sqrt{3}}
      (t-\Delta_{K \pi} )\,\bigg ] \overline{B}_{21}(M_\pi,M_\pi;t) \nonumber \\
  &-& 2t\bigg [\,3(s-u)- \Delta _{\pi}-\frac{\epsilon}{\sqrt{3}}
      (t-4 \Delta_{K \pi} )\,\bigg ] \overline{B}_{22}(\Mpic,\Mpin;t) \,\bigg\}\,. 
\end{eqnarray}
\begin{eqnarray}
{\cal M}_{\pi^-\eta}
  &=& \frac{1}{3\sqrt{2}F_0^4}\bigg (\,\frac{\epsilon}{\sqrt{3}}
        \,\bigg )\bigg\{\,-2
       (4\MK -3t)F_0^2\mueta 
        +(\Sigma_{\pi \eta} -t)^2 \overline{B}(\Mpi,\Meta^2;t) \nonumber \\
  &-& 2t(\Sigma_{\pi \eta} -t) \overline{B}_1(\Mpi,\Meta^2;t)
 +t^2 \overline{B}_{21}(\Mpi,\Meta^2;t) +t^2 \overline{B}_{22}(\Mpi,\Meta^2;t)
 \,\bigg\}\,. \nonumber \\ 
\end{eqnarray}
\begin{eqnarray}
&&{\cal M}_{K^-{\bar K}^0}
  = -\frac{1}{12\sqrt{2}F_0^4}\bigg\{\,-\frac{4\epsilon}{\sqrt{3}}
       (3t-2\MK )F_0^2\mu_K \nonumber \\
  &+& 2t\bigg [\,\Delta _{\pi}+\frac{\epsilon}{\sqrt{3}}
       (2\MK -t)\,\bigg ] \overline{B}(M_K,M_K;t) 
 +\frac{4\epsilon}{\sqrt{3}}\,t(\MK -t)
 \overline{B}_1(M_K,M_K;t) \nonumber \\
  &-& 2t(\Delta _{\pi}+\frac{\epsilon}{\sqrt{3}}\,
      t)  \overline{B}_{21}(M_K,M_K;t)
      -2t\bigg [\,3(s-u)+\Delta _{\pi}+\frac{\epsilon}{\sqrt{3}}\,
      t\,\bigg ]  \overline{B}_{22}(\MKc,\MKn;t)\,\bigg\}\,.\nonumber \\ 
\end{eqnarray}

Finally for the u-channel we get the following set of results. 
\begin{equation}
{\cal M}\Bigm|_{\textrm{u-chanel}} = {\cal M}_{K^-\pi^0}+
{\cal M}_{K^-\eta}+{\cal M}_{\pi^-K^0}\,.
\end{equation}
{F}or each term separately
\begin{eqnarray}
&&{\cal M}_{K^-\pi^0}=
-\frac{1}{24\sqrt{2}F_0^4}\bigg ( 
\nonumber\\
&-&2F_0^2\bigg [\,5\Mpin +3\MKn -3u+3\Delta _{\pi}
      -\frac{\epsilon}{\sqrt{3}}(15u-29\Mpin +13\MKn )
       \,\bigg ] \mupin \nonumber \\           
 &+& (1-\frac{\epsilon}{\sqrt{3}})
          (\Sigma _{K^-\pi^+}-u+
           \Delta _{\pi})\bigg [\,5\MKc +\Mpin -u
           +\frac{6\epsilon}{\sqrt{3}}
 (\Sigma _{\pi K}-u)\,\bigg ] \overline{B}(\MKc,\Mpin;u) \nonumber \\
 &+& \left\{-\bigg [\,5\MKc +\Mpin -u
           +\frac{6\epsilon}{\sqrt{3}}
           (\Sigma _{\pi K}-u)\,\bigg ]\bigg [\,\Delta _{\pi^+K^0}
           +u+\frac{\epsilon}{\sqrt{3}}(3\Delta _{\pi K}-u)
           \,\bigg ] \right. \nonumber \\
  &+& \left. (1-\frac{\epsilon}{\sqrt{3}})
 (1+\frac{6\epsilon}{\sqrt{3}})(\Sigma _{K^-\pi^+}-u+
 \Delta _{\pi})(3\Delta _{\pi^0K^-}-u) \right\} 
 \overline{B}_1(\MKc,\Mpin;u) \nonumber \\
&-& (1+\frac{6\epsilon}{\sqrt{3}})(3\Delta _{\pi^0K^-}-u)
          \bigg [\,
           \Delta _{\pi^+K^0}+u+\frac{\epsilon}{\sqrt{3}}
           (3\Delta _{\pi K}-u)
          \,\bigg ] \overline{B}_{21}(\MKc,\Mpin;u) \nonumber \\
&-& 2u(1+\frac{6\epsilon}{\sqrt{3}})
          \bigg [\,\Sigma _{\pi}-t 
      + 2(s-\Sigma _{\pi^+K^-}) \nonumber \\
& +&\frac{\epsilon}{\sqrt{3}}(2\Mpi -t)+
          \frac{4\epsilon}{\sqrt{3}}(2\MK -t)-
\frac{4\epsilon}{\sqrt{3}}(\Sigma _{\pi K}-s)\,\bigg ] 
\overline{B}_{22}(\MKc,\Mpin;u)
\bigg)\,.
\end{eqnarray}
\begin{eqnarray}
{\cal M}_{K^-\eta}
 &=& \frac{1}{24\sqrt{2}F_0^4}\left(1+\frac{2\epsilon}{\sqrt{3}}\right)
 \bigg (\, \nonumber \\ 
      &-& 2F_0^2\bigg [\,5\Mpin -\MKn -3u+3\Delta _{\pi}
          -\frac{3\epsilon}{\sqrt{3}}(3u-\Mpi -3\MK )
          \,\bigg ] \mueta \nonumber \\           
&+& (1+\frac{3\epsilon}{\sqrt{3}})
          (\Sigma _{K^-\eta}-u-2
           \Delta _{\pi})(\Sigma _{K^-\eta}-u+4
           \Delta _{\pi}) \overline{B}(\MKc,\Meta^2;u) \nonumber \\
&+& \left\{ (\Sigma _{K^-\eta}-u+4
           \Delta _{\pi})\bigg [\,
           3\Delta _{\pi^+K^0}-u-\frac{3\epsilon}{\sqrt{3}}
           (\Delta _{\pi K}+u)\,\bigg ] \right .\nonumber \\
 &+& \left. (1+\frac{3\epsilon}{\sqrt{3}})(\Sigma _{K^-\eta}-u-2
\Delta _{\pi})(3\Delta _{\pi^0K^-}-u) \right\} 
\overline{B}_1(\MKc,\Meta^2;u) \nonumber \\
&+& (3\Delta _{\pi^0K^-}-u)
      \bigg [\,
      3\Delta _{\pi^+K^0}-u-\frac{3\epsilon}{\sqrt{3}}
         (\Delta _{\pi K}+u)
        \,\bigg ] \overline{B}_{21}(\MKc,\Meta^2;u) \nonumber \\
&+& 2u\bigg [\,2(s-t-u) 
+ \Sigma _{\pi}-t+4(\Sigma _K-t) \nonumber \\
 &+&\frac{6\epsilon}{\sqrt{3}}(\Sigma _{\pi K}-s)
-\frac{3\epsilon}{\sqrt{3}}(2\Mpi -t)\,\bigg ]  
\overline{B}_{22}(\MKc,\Meta^2;u)\bigg)\,.
\end{eqnarray}
\begin{eqnarray}
{\cal M}_{\pi^-K^0}
      &=& -\frac{1}{12\sqrt{2}F_0^4}\bigg ( 
-2F_0^2(1-\frac{\epsilon}{\sqrt{3}})(3u-3\Mpin +\MKn -3
        \Delta _{\pi}) \muKn \nonumber  \\           
&+& (1-\frac{\epsilon}{\sqrt{3}})
           (u+\Delta _{\pi^-K^0})(\Sigma _{\pi^-K^0}-u+2
    \Delta _{\pi})  \overline{B}(\Mpic,\MKn;u) \nonumber \\
  &+& \left\{ -(1-\frac{\epsilon}{\sqrt{3}})
           (\Sigma _{\pi^-K^0}-u+2
           \Delta _{\pi})(3\Delta _{\pi^-K^0}-u) \right. \nonumber \\
      &+& \left. (u+\Delta _{\pi^-K^0})\bigg [\,
   \Delta _{\pi^0K^-}-u+\frac{\epsilon}{\sqrt{3}}(u+3\Delta _{\pi K})\,
     \bigg ] \right\}\overline{B}_{1}(\Mpic,\MKn;u) \nonumber \\
      &-& (3\Delta _{\pi^-K^0}-u)\bigg [\,\Delta _{\pi^0K^-}-u+
          \frac{\epsilon}{\sqrt{3}}(u+3\Delta _{\pi K})
          \,
           \bigg ]  \overline{B}_{21}(\Mpic,\MKn;u) \nonumber \\
      &-& 2u\bigg [\,s-\Sigma _{\pi^+K^-}+2(\Sigma _K-t)
           +\frac{4\epsilon}{\sqrt{3}}(\Sigma _{\pi K}-t)-
 \frac{5\epsilon}{\sqrt{3}}(\Sigma _{\pi K}-s)\,\bigg ] \times 
 \nonumber \\&& \hspace{1cm} \overline{B}_{22}(\Mpic,\MKn;u)
 \bigg)\,. \nonumber\\
\end{eqnarray}

The only remaining piece is the soft 
photon exchange diagram. 
It contains an infrared divergence (included in the
$C$ function). 
Its result is given by
\begin{eqnarray}
{\cal M}\Bigm|_{\textrm{one-photon}}
   &=& -\frac{e^2}{4\sqrt{2}F_0^2}(u+\Delta _{\pi K})
        \bigg (\,-6F_0^2\mupitild +\frac{1}{4\pi ^2}\,\bigg ) \nonumber\\
   &-& \frac{e^2}{4\sqrt{2}F_0^2}(u-\Delta _{\pi K})
        \bigg (\,-6F_0^2\muKtild +\frac{1}{4\pi ^2}\,\bigg ) \nonumber \\
   &+& \frac{e^2}{2\sqrt{2}F_0^2}\bigg\{\,
        -F_0^2\mupitild (\Sigma _{\pi K}+u-2s)
        -F_0^2\muKtild (\Delta _{\pi K}+u-2s) \nonumber \\
   &+& 2(u-s)\bigg (\,\frac{1}{16\pi ^2}\,\bigg )
        +(\Sigma _{\pi K}-s)\overline{B}(M_\pi^2 M_K^2;s)
        +(s-\Delta _{\pi K}) \overline{B}_1(M_\pi^2 M_K^2;s) \nonumber\\
   &+& 2(s-u)(s-\Sigma _{\pi K})C_{\pi K}(s)
        +4(s-\Sigma _{\pi K})\bigg [\,uG_{\pi K}^-(s)+
\Delta _{\pi K}G_{\pi K}^+(s)\,\bigg ]\,\bigg\} \,, \nonumber \label{eq:one photon exchange contribution} \\
\end{eqnarray}
where the functions $G_{\pi K}^-$ an $G_{\pi K}^+$ are defined in the
following appendix.


\subsection{Scattering lengths}
\label{scl}

Due to the relevance of the process $\pi^- K^+ \rightarrow \pi^0 K^0$
we display the combination for the
S-wave scattering lengths.
After performing
all the steps mentioned in sec.~\ref{sec:TH} we make the following
substitutions
\[
\Delta_K \rightarrow \Delta_\pi -\frac{4\epsilon}{\sqrt{3}}(M_K^2
-M_\pi^2)\,,\quad \Delta_\pi \rightarrow 2e^2 Z F_0^2\,.
\]
This leads to the following expression
\begin{eqnarray}
&&- \frac{3}{\sqrt{2}}a_0(+-;00) = 
a^{1/2}_0 - a^{3/2}_0 - \frac{3}{\sqrt{2}} \Delta_0(+-;00) = \\ \no && 
\frac{3}{32\pi} \frac{M_{\pi^\pm} M_{K^\pm}}{F_0^2}
+\frac{3}{256\pi F_0^2}\frac{M_K-M_\pi}{M_K+M_\pi}\Delta_K + \frac{3}{256\pi F_0^2}
\frac{3M_K+5M_\pi}{M_K+M_\pi}\Delta_\pi+\frac{\sqrt{3}}{64\pi F_0^2}(M^2_K+M_\pi^2)
\epsilon
\\ \no && 
+
\frac{3}{32\pi}
\frac{M_{\pi^\pm} M_{K^\pm}}{F_0^4}
\left\{  
-8
\left( M_{\pi^\pm}^2 + 2 M_{K^\pm}^2 \right) L_4^r - \frac{1}{576 \pi^2 } 
\left(\frac{1}{M_{K^\pm}^2-M_{\pi^\pm}^2}\right) \right. 
\\ \no && ~~
\left[27 M_{\pi^\pm}^4 \log\left(\frac{M_{\pi^\pm}^2}{\mu^2}\right) -2 M_{K^\pm}^2
(18 M_{K^\pm}^2 + 5 M_{\pi^\pm}^2) \log\left(\frac{M_{K^\pm}^2}{\mu^2}\right) 
\right.
\\ \no && ~~
\left. \left.
+M_{\pi^\pm}^2 (28 M_{K^\pm}^2 - 9 M_{\pi^\pm}^2) \log\left(\frac{M_{\eta}^2}{\mu^2}\right)
\right]
\right\} 
+ \frac{M_{\pi^\pm}^2 M_{K^\pm}^2}{32\pi F_0^4} 
{\cal B}_- 
\\ \no && 
-\frac{\sqrt{3}\epsilon}{32 \pi F^4_0} 
\left\{  
-\frac{M_\pi}{432 \pi^2 (M_K-M_\pi) (4M_K^2-M_\pi^2)} \times
\right.
\\ \no && 
\left(216 M_K^6-604M_K^5M_\pi+862M_K^4M_\pi^2 +1331 M_K^3 M_\pi^3-245 M_K^2 M_\pi^4 
-283M_KM_\pi^5 +M_\pi^6\right)
\\ \no && 
+
4\left[ (M_K^4 -M_\pi^4) (L_5^r -2 L_8^r) +2 M_K M_\pi (2 L_3^r (M_K-M_\pi)
M_\pi+L_4^r(6M_K^2-3M_\pi^2) ) \right]
\\ \no && 
+\frac{M_KM_\pi^2}{36(M_K^2-M_\pi^2)(4M_K^2-M_\pi^2)} \times
\\ \no && 
\left[ \frac{}{}\right.
\left( 
304M_K^5-60M_K^4M_\pi-472M_K^3M_\pi^2+10M_K^2M_\pi^3+99M_KM_\pi^4+8M_\pi^5\right)
 {\cal B}_+ 
\\ \no && 
+
\left.
\left( 
-160M_K^5+24M_K^4M_\pi+310M_K^3M_\pi^2+26M_K^2M_\pi^3-81M_KM_\pi^4-8M_\pi^5\right)
 {\cal B}_-\frac{}{} \right]
\\ \no && 
+
\frac{1}{ \pi^2 (M_K-M_\pi)^2 (M_K+M_\pi)} \times
\\ \no && 
\left[ \frac{}{}\right.
\frac{M_\pi^2}{ 192 }
(71 M_K^5-79M_K^4 M_\pi -174M_K^3 M_\pi^2 +43 M_K^2 M_\pi^3 -44 M_K M_\pi^4 + 3 M_\pi^5)
\log\left(\frac{M_\pi^2}{\mu^2}\right)
\\ \no && 
+
\frac{M_K M_\pi}{ 864 } 
(-162M_K^5+68M_K^4M_\pi+455M_K^3M_\pi^2+645M_K^2M_\pi^3-154M_KM_\pi^4+128M_\pi^5)
\log\left(\frac{M_K^2}{\mu^2}\right)
\\ \no && 
+
\frac{1}{1728 } 
(-36M_K^7+36M_K^6M_\pi-415M_K^5M_\pi^2+251M_K^4M_\pi^3
\\ \no && ~~
\hspace{4.5cm}
\left. \left.
-174M_K^3M_\pi^4
-277M_K^2M_\pi^5+266M_KM_\pi^6
+9M_\pi^7)
\log\left(\frac{M_\eta^2}{\mu^2}\right) \right] \frac{}{}\right\}
\\ \no && 
-
\frac{3Ze^2}{32\pi F_0^2}\left\{\frac{}{}
\frac{1}{1728\pi^2(M_K-M_\pi)^2(4M_K^2-M_\pi^2)(M_K+M_\pi)} \times
\left(
4M_K^7-124M_K^6M_\pi+671M_K^5M_\pi^2
\right. \right.
\\ \no && 
\hspace{4cm}
\left.
+3967M_K^4M_\pi^3+2276M_K^3M_\pi^4
-1136M_K^2M_\pi^5-575M_KM_\pi^6+29M_\pi^7 \right)
\\ \no && 
+
2\left[ (8L_4^r+3L_5^r)M_K^2-4(L_3^r+6L_4^r)M_KM_\pi
+(4L_4^r-L_5^r)M_\pi^2\right]
\\ \no && 
+\frac{M_K M_\pi}{72(M_K-M_\pi)^2(M_K+M_\pi)(4M_K^2-M_\pi^2)}\times
\\ \no && 
\left[ \frac{}{}
\right.
(-160M_K^5+48M_K^4M_\pi+274M_K^3M_\pi^2+14M_K^2M_\pi^3-45M_KM_\pi^4-20M_\pi^5)
{\cal B}_+
\\ \no && 
+
(-32M_K^5+252M_K^4M_\pi-88M_K^3M_\pi^2-362M_K^2M_\pi^3+51M_KM_\pi^4+68M_\pi^5)
{\cal B}_- \left. \frac{}{}\right]
\\ \no && + 
\frac{1}{\pi^2(M_K-M_\pi)^2(M_K^2-M_\pi^2)}\times
\\ \no && 
\left[\frac{}{} \right.
\frac{M_\pi}{192}\left(8M_K^5-4M_K^4M_\pi+30M_K^3M_\pi^2+70M_K^2M_\pi^3
-29M_K M_\pi^4+15M_\pi^5\right)
\log\left(\frac{M_\pi^2}{\mu^2}\right)
\\ \no && -
\frac{M_K}{1728}\left(27M_K^5-220M_K^4M_\pi+854M_K^3M_\pi^2-54M_K^2M_\pi^3
+269M_KM_\pi^4+104M_\pi^5\right)\log\left(\frac{M_K^2}{\mu^2}\right)
\\ \no && 
+\left.
\frac{M_\pi}{1728}\left(32M_K^5+296M_K^4M_\pi-270M_K^3M_\pi^2+260M_K^2M_\pi^3-121M_KM_\pi^4-27M_\pi^5
\right)\log\left(\frac{M_\eta^2}{\mu^2}\right) \right]
\left. \frac{}{}\right\} 
\\ \no && 
-\frac{3 e^2}{64 \pi^3F_0^2}\left\{
\frac{M_\pi^2 \pi^2 }{9 } \left(\frac{2 M_K^2+M_\pi^2}{M_K^2-M_\pi^2}
\right) \left[ 6 K_3^r-3K_4^r-2\left(K_5^r+K_6^r-K_9^r-K_{10}^r\right)\right]
\right.
\\ \no && +
\frac{ M_K^2}{64 } \left[ -12 +\left(\frac{9 M_K+11 M_\pi}{M_K+M_\pi} \right)
\log\left(\frac{M_K^2}{\mu^2}\right) + 64 \pi^2 [K_5^r+K_6^r-6(K_{10}^r+K_{11}^r) ] \right]
\\ \no && +
\frac{ M_\pi^2}{144 } \left[ 9 -\frac{9}{4} \left(\frac{M_K+3 M_\pi}{M_K+M_\pi}\right) 
\log\left(\frac{M_\pi^2}{\mu^2}\right) \right.
\\ \no && 
-8\pi^2 \left( 6K_3^r-3K_4^r+4K_5^r+4K_6^r+2K_9^r-34K_{10}^r-36K_{11}^r\right)\left. \frac{}{}\right]
\\ \no && +
\frac{ M_\pi M_K}{36} \left[\frac{}{} 8 \pi^2 \left(24 K_1^r+24 K_2^r+18K_3^r-9K_4^r+20 K_5^r + 2 K_6^r\right)
\right. \\ \no && 
+\frac{9M_K^2}{2(M_K+M_\pi)^2} 
\left( 2 + \log\left(\frac{M_K^2}{\mu^2}\right) 
+\log\left(\frac{M_\pi^2}{\mu^2}\right) \right) 
\\ \no && 
+\left. \left. 
\frac{9M_\pi^2}{2(M_K+M_\pi)^2} \left(10+3 \log\left(\frac{M_K^2}{\mu^2}
\right) - \log\left(\frac{M_\pi^2}{\mu^2}\right)\right) 
+\frac{18 M_K M_\pi}{(M_K+M_\pi)^2} 
\left(3+\log\left(\frac{M_\pi^2}{\mu^2}\right)\right) \right]
\right\}\,,
\end{eqnarray}
where for sake of clarity we have 
expressed the combination $a_0(+-;00)$ in terms of 
the bared coupling 
constant $F_0$. 
If one chooses to renormalize the decay constants as $F_\pi F_K$
the following term should be added to the previous expression
\begin{eqnarray}
&& \delta_{F_{\pi}F_K}\,=  
-
\frac{3M_{\pi^{\pm}}M_{K^{\pm}}}{256 \pi}\left(\frac{1}{16\pi^2F_0^4}\right)
\bigg\{-512\pi^2[2(\Mpic +2\MKc )L_4^r
     +(\Mpic +\MKc )L_5^r] \nonumber \\
&& +11\Mpic\log\left (\frac{\Mpic}{\mu^2}\right )+10\MKc\log\left (\frac{\MKc}{\mu^2}\right )
     +(4\MKc -\Mpic )\log\left (\frac{\Meta^2}{\mu^2}\right )\bigg\} \nonumber \\
&& -\frac{3 M_{\pi}M_K}{256\pi}\left(\frac{1}{16\pi^2F_0^4}\right)
\left (\frac{\epsilon}{\sqrt{3}}\right )
     \bigg\{20(\MK -\Mpi ) 
 -512\pi^2[3\MK (4L_4^r+L_5^r)-\Mpi (6L_4^r+L_5^r)]  \nonumber \\
&& +11\Mpi\log\left (\frac{\Mpi}{\mu^2}\right )+10(3\MK -2\Mpi )
\log\left (\frac{\MK}{\mu^2}\right )+3(4\MK -3\Mpi )\log\left (\frac{\Meta^2}{\mu^2}\right )
\bigg\} \nonumber \\
&& +\frac{3 Z e^2}{256\pi}\left(\frac{1}{16 \pi^2F_0^2}\right) \bigg\{42M_{\pi}M_K 
+512\pi^2[\MK (4L_4^r+L_5^r)-4M_{\pi}M_K(3L_4^r+L_5^r)+\Mpi (2L_4^r+L_5^r)] \nonumber \\
&& +11M_{\pi}(2M_K-M_{\pi})\log\left (\frac{\Mpi}{\mu^2}\right )-10M_K(M_K-2M_{\pi})
     \log\left (\frac{\MK}{\mu^2}\right )
     \\ \no &&
     +(-4\MK +6M_{\pi}M_K+\Mpi )
     \log\left (\frac{\Meta^2}{\mu^2}\right )\bigg\}\,. \no \\ 
\end{eqnarray} 
Otherwise, if the desired renormalization is as $F_\pi^2$ the term to
be added is
\begin{eqnarray}
&& \delta_{F_{\pi}^2}\,=  
\frac{3 M_{\pi^\pm}M_{K^\pm}}{32\pi}
\left(\frac{1}{16\pi^2F_0^4}\right)\times
\\ \no&&
\bigg\{128\pi^2[(\Mpic +2\MKc )L_4^r
  +\Mpic L_5^r] 
 -2\Mpic\log\left (\frac{\Mpic}{\mu^2}\right )-\MKc
 \log\left (\frac{\MKc}{\mu^2}\right )\bigg\} \nonumber \\
&& -\frac{3M_{\pi}M_K}{32\pi}\left(\frac{1}{16\pi^2F_0^4}\right)
\left (\frac{\epsilon}{\sqrt{3}}\right )
     \bigg\{2(\MK -\Mpi ) 
-128\pi^2[3(2\MK -\Mpi )L_4^r+\Mpi L_5^r] \nonumber \\
&& +2\Mpi\log\left (\frac{\Mpi}{\mu^2}\right )+(3\MK -2\Mpi )\log\left (\frac{\MK}{\mu^2}\right )\bigg\} \\ \no
&&
-\frac{3Ze^2}{32\pi}\left(\frac{1}{16\pi^2F_0^2}\right)\bigg\{-6M_{\pi}M_K 
-128\pi^2[2\MK L_4^r-2M_{\pi}M_K(3L_4^r+L_5^r)+\Mpi (L_4^r+L_5^r)] \nonumber \\
&& +2M_{\pi}(-2M_K+M_{\pi})\log\left (\frac{\Mpi}{\mu^2}\right )+M_K(M_K-2M_{\pi})
 \log\left (\frac{\MK}{\mu^2}\right )\bigg\}\,. \nonumber \\ 
\end{eqnarray} 
Obviously once the renormalization of the coupling constant is taken
into account there are some cancellations that simplify slightly the expression
for the combination of scattering lengths.
In writing this expression we have applied extensively the Gell-Mann--Okubo
relation to the polynomial terms 
and to the $\bar{J}$ functions but otherwise
keep the eta mass inside the $\log$ functions. The functions ${\cal B}_+$
and ${\cal B}_-$
are defined in the next appendix.
Notice that this combination of scattering lengths 
is independent at leading order on the ratio $\frac{m_s}{\hat{m}}$, this 
implies that the uncertainty from this quantity 
is very small \cite{KSSF}.

\setcounter{section}{0}
\setcounter{subsection}{0}

\renewcommand{\thesection}{\Alph{zahler}}
\renewcommand{\theequation}{\Alph{zahler}.\arabic{equation}}

\setcounter{equation}{0}
\addtocounter{zahler}{1}
\renewcommand{\thesection}{\Alph{zahler}}
\renewcommand{\theequation}{\Alph{zahler}.\arabic{equation}}

\section{Loop integrals}
\label{int}

In this appendix we collect 
for completeness some familiar formulas.
The $\overline{B}_{ij}$ functions are the finite components of those defined
in \cite{pasarino}.
Using Lorenz decomposition and some simpler algebraic manipulation
they 
can be reduced to the 
tadpole integral, $\mu$, 
and the one-loop function, $\overline{J}$, through the following
set of finite
relations 
\begin{eqnarray}
&&\overline{B}(m_1^2,m_2^2;p^2) =
\frac{\mu_{m_1}-\mu_{m_2}}{\Delta_{m_1m_2}} + \overline{J}(m
_1^2,m_2^2;p^2)\,,
\nonumber \\
&&\overline{B}_1(m_1^2,m_2^2;p^2)\,=\,\frac{1}{2}\left [\left (1+\frac{\Delta_{m_1m_2}}{p^2}\right )
\overline{J}(m_1^2,m_2^2;p^2)+\frac{\mu_{m_1}-\mu_{m_2}}{\Delta_{m_1m_2}}\right ]\,, \nonumber \\
\nonumber \\
&&\overline{B}_{21}(m_1^2,m_2^2;p^2)\,=\,\frac{1}{3p^2}\bigg\{\frac{1}{p^2}\left [\lambda_{m_1m_2}(p^2)+3p^2m_1^2\right ]
\overline{J}(m_1^2,m_2^2;p^2) \nonumber \\
&&+\left (\frac{p^2-m_2^2}{\Delta_{m_1m_2}}\right )\mu_{m_1}-\left (\frac{p^2-m_1^2}
{\Delta_{m_1m_2}}\right )\mu_{m_2}+\frac{p^2}{6}\left (\frac{1}{16\pi^2}\right )\left (1-\frac{3\Sigma_{m_1m_2}}{p^2}\right )
\bigg\}\,, \nonumber \\
\nonumber \\
&&\overline{B}_{22}(m_1^2,m_2^2;p^2)\,=\,-\frac{1}{12p^4}\bigg\{\lambda_{m_1m_2}(p^2)\overline{J}(m_1^2,m_2^2;p^2)+
\frac{2p^4}{3}\left (\frac{1}{16\pi^2}\right )\left (1-\frac{3\Sigma_{m_1m_2}}{p^2}\right ) \nonumber \\
&&-\left (p^2+\Delta_{m_1m_2}\right )\mu_{m_1}-\left (p^2-\Delta_{m_1m_2}\right )\mu_{m_2}+\lambda_{m_1m_2}(p^2)\,
\frac{\mu_{m_1}-\mu_{m_2}}{\Delta_{m_1m_2}}
\bigg\}\,. \nonumber \\
\end{eqnarray}
In the  $\overline{J}$ function we have to consider at least two branches 
\beqa
32 \pi^2 \overline{J}(m_1^2,m_2^2;p^2) = \left\{ \begin{array}{l}
{\bf i)}\, \mbox{if}\quad  p^2 \ge (m_1+m_2)^2\,.   \vspace{0.05in} \\ 
\hspace{0.5cm}2+\left(-\frac{\Delta_{m_1m_2}}{p^2}+\frac{\Sigma_{m_1m_2}}{\Delta_{m_1m_2}}
\right)\log\left(\frac{m_1^2}{m_2^2}\right)
-\frac{\lambda_{m_1m_2}}{p^2}\log\frac{(p^2+\lambda_{m_1m_2})^2-\Delta_{m_1m_2}^2}
{(p^2-\lambda_{m_1m_2})^2-\Delta_{m_1m_2}^2}\,.   \vspace{0.05in} \\ 
{\bf ii)}\, \mbox{if}\quad (m_1-m_2)^2 \le p^2 \le (m_1+m_2)^2\,.\vspace{0.05in} \\
\hspace{0.5cm}2+\left(-\frac{\Delta_{m_1m_2}}{p^2}+\frac{\Sigma_{m_1m_2}}{\Delta_{m_1m_2}}
\right)\log\left(\frac{m_1^2}{m_2^2}\right) \vspace{0.05in} \\ 
\hspace{0.5cm}-2\,\frac{\sqrt{-\lambda_{m_1m_2}^{1/2}}}{p^2}\left [ 
\mbox{arctg}\left(\frac{p^2+\Delta_{m_1m_2}}{\sqrt{-\lambda_{m_1m_2}^{1/2}}}\right) 
- \mbox{arctg}\left(\frac{-p^2+\Delta_{m_1m_2}}{\sqrt{-\lambda_{m_1m_2}^{1/2}}}\right)\right ]\,.
\end{array}
\right.
\eeqa

With the use of the Gell-Mann--Okubo relation the $\overline{J}$ function 
can be reduced
to simpler combinations of $\log$ and arctan functions. In the 
latter case we 
define following the functions
\[
{\cal B}_\pm \equiv {\cal B}(M_K,M_\pi)\pm {\cal B}(M_K,-M_\pi)\,, 
\]
where
\begin{eqnarray}
{\cal B}(x,y) = - \frac{\sqrt{(x-y)(2x+y)}}{12\pi^2 (x+y)}
\left[ \mbox{arctg}\left(\frac{\sqrt{(x-y)(2x+y)}}{2 (x-y)}\right)
+ \mbox{arctg}\left(\frac{ x+2y}{2\sqrt{(x-y)(2x+y)}}\right) \right]\,.
\end{eqnarray}

Besides the previous integrals in the calculation
is needed the three-point function integrals. Those arise through the photon exchange
diagram.
The scalar function, the only one that is actually IR divergent, is 
given in eq.~(\ref{Co}) while the remaining one is
\begin{eqnarray}
C^{\mu }(m_P^2,m_Q^2,m_\gamma^2;p,k)
  &\equiv & -i\mu^{4-D}\int\frac{d^Dl}{(2\pi )^D}\,\frac{l^{\mu }}
              {(l^2-m_{\gamma }^2)[(p-l)^2-M_P^2][(k-l)^2-M_Q^2]} \nonumber \\
&= & (p-k)^{\mu }G_{PQ}^-(p,k)+(p+k)^{\mu }G_{PQ}^+(p,k)\,.
\end{eqnarray}
In the convention of \cite{pasarino} the previous decomposition reads
\[
G^+_{PQ}(p,k) \propto -\frac{C_{11}(p,k)}{2}\,, \quad G^-_{PQ}(p,k) \propto -\frac{C_{11}(p,k)}{2}+C_{12}(p,k)\,.
\]
In terms of the basic functions we obtain
\begin{eqnarray}
G_{PQ}^-(q^2,p\cdot k)
  &=& \frac{1}{{\cal G}}\left\{-\Delta _{PQ}\bigg [{\bar J}_{PQ}(q^2)
       -\frac{1}{16\pi^2}\bigg ] 
  \quad -\frac{1}{32\pi^2}[(p+k)^2-\Sigma _{PQ}]
        \log\left(\frac{M_P^2}{M_Q^2}\right) \right\}\,,
         \no \\
G_{PQ}^+(q^2,p\cdot k)&=&\frac{1}{{\cal G}}\left\{q^2\bigg [{\bar J}_{PQ}(q^2)-
       \frac{1}{16\pi^2}\bigg ]
       -\frac{1}{32\pi^2}\bigg (q^2\,\frac{\Sigma _{PQ}}{\Delta _{PQ}}-
       \Delta _{PQ}\bigg ) \log\left(\frac{M_P^2}{M_Q^2}\right)
       \right\}\,,
\end{eqnarray}
where
\[
{\cal G} = [q^2(p+k)^2-\Delta _{PQ}^2]\,~ \mbox{and}~ q^2 = (p-k)^2\,.
\]


\end{document}